\documentclass[preprint,aps,pre]{revtex4}
\usepackage{amsmath,amssymb}
\usepackage{epsfig}
\usepackage{graphics}
\usepackage{graphicx}
\usepackage{float}
\usepackage{dcolumn}

\begin{document}

\title{RESEARCH METHODOLOGY}

\author{S.~Rajasekar}

\affiliation{School of Physics, Bharathidasan University,
Tiruchirapalli -- 620 024, Tamilnadu, India}
\email{rajasekar@cnld.bdu.ac.in}

\author{P.~Philominathan}

\affiliation{Department of Physics, Sri AVVM Pushpam College,
Poondi, Thanjavur -- 613 503, Tamilnadu, India}

\author{V.~Chinnathambi}

\affiliation{Department of Physics, AKGS Arts College,
Srivaikundam -- 628 619, Tamilnadu, India}

%\date{}

\begin{abstract}
In this manuscript various components of research are listed and
briefly discussed.  The topics considered in this write-up cover a
part of the research methodology paper of Master of Philosophy
(M.Phil.) course and Doctor of Philosophy (Ph.D.) course.  The
manuscript is intended for students and research scholars of
science subjects such as mathematics, physics, chemistry,
statistics, biology and computer science.  Various stages of
research are discussed in detail.  Special care has been taken to
motivate the young researchers to take up challenging problems.
Ten assignment works are given.  For the benefit of young
researchers a short interview with three eminent scientists is
included at the end of the manuscript.
\end{abstract}

\maketitle
\section{What is Research?}

{\emph{Research}} is a logical and systematic search for new and useful
information on a particular topic.  In the well-known nursery rhyme
 \vskip 5pt
{\emph{Twinkle Twinkle Little Star }}

 \vskip 2pt
{\emph{How I Wonder What You Are}}
 \vskip 5pt
\noindent the use of the words {\emph{how}} and {\emph{what}} essentially
summarizes what research is.  It is an investigation of
finding solutions to scientific and social problems through
objective and systematic analysis.  It is a search for knowledge,
that is, a discovery of hidden truths. Here knowledge means
information about matters. The information might be collected from
different sources like experience, human beings, books, journals,
nature, etc.  A research can lead to new contributions to the
existing knowledge. Only through research is it possible to make
progress in a field.  Research is indeed civilization and determines
the economic, social and political development of a nation.
The results of scientific research very often force a change in the
philosophical view of problems which extend far beyond the
restricted domain of science itself.
 \vskip 10pt
Research is not confined to science and technology only.  
There are vast areas of research in other disciplines such as languages,
literature, history and sociology.  Whatever might be the subject, 
research has to be an active, diligent and systematic process of 
inquiry in order to discover, interpret or revise facts, events, 
behaviours and theories.  Applying the outcome of research for the
refinement of knowledge in other subjects, or in enhancing the 
quality of human life also becomes  a kind of research and development.
 \vskip 10pt
Research is done with the help of study, experiment, observation,
analysis, comparison and reasoning. Research is in fact
ubiquitous.  For example, we know that cigarette smoking is
injurious to health; heroine is addictive; cow dung is a useful
source of biogas; malaria is due to the virus protozoan
plasmodium; AIDS (Acquired Immuno Deficiency Syndrome) is due to
the virus HIV (Human Immuno Deficiency Virus).  How did we know
all these? We became aware of all these information only through
research.  More precisely, it seeks predictions of events,
explanations, relationships and theories for them.
 \vskip 10pt
As stated by Gerald Milburn
%Quantum Technology 1996 Allen & Unwin, Sydney
{\emph{Scientific research is a chaotic business, stumbling along
amidst red herrings, errors and truly, creative insights. Great
scientific breakthroughs are rarely the work of a single
researchers plodding slowly by inexorably towards some final goal.
The crucial idea behind the breakthrough may surface a number of
times, in different places, only to sink again beneath the babble
of an endless scientific discourse}}.
\subsection{\large{\bf{\emph{What are the Objectives of
Research?}}}}
The prime objectives of research are
\renewcommand{\labelenumi}{(\theenumi)}
\renewcommand{\theenumi}{\arabic{enumi}}
\begin{enumerate}
 \item
to discover new facts
 \item
to verify and test important facts
 \item
to analyse an event or process or phenomenon to identify the cause
and effect relationship
 \item
to develop new scientific tools, concepts and theories to solve
and understand scientific and nonscientific problems
 \item
to find solutions to scientific, nonscientific and social problems
and
 \item
to overcome or solve the problems occurring in our every day life.
\end{enumerate}
\subsection{\large{\bf{\emph{What Makes People do Research?}}}}
This is a fundamentally important question.  {\emph{No person
would like to do research unless there are some motivating
factors}}. Some of the motivations are the following:
\renewcommand{\labelenumi}{(\theenumi)}
\renewcommand{\theenumi}{\arabic{enumi}}
\begin{enumerate}
 \item
to get a research degree (Doctor of Philosophy (Ph.D.)) along with
its benefits like better employment, promotion, increment in
salary, etc.
 \item
to get a research degree and then to get a teaching position in a
college or university or become a scientist in a research
institution
 \item
to get a research position in countries like U.S.A., Canada,
Germany, England, Japan, Australia, etc. and settle there
 \item
to solve the unsolved and challenging problems
 \item
to get joy of doing some creative work
 \item
to acquire respectability
 \item
to get recognition
 \item
curiosity to find out the unknown facts of an event
 \item
curiosity to find new things
 \item
to serve the society by solving social problems.
\end{enumerate}
\noindent Some students undertake research without any aim
possibly because of not being able to think of anything else to
do.  Such students can also become good researchers by motivating
themselves toward a respectable goal.  As pointed out by Prof. Rajesh Kasturirangan (NIAS, IISc) even if you work in a company or run a company, a mind inclined towards research would do better than a mind  not trained for it and it was like the story of the hare and the tortoise.  If you have a mind trained for research, you will be the tortoise -- the climb would be slow and steady, but eventually you would win the race.
\subsection{\large{\bf{\emph{Thesis Research}}}}
In the words of Prof.P.~Balaram [Current Science, 87(2004)1319]
{\emph{Ph.D. degree is a passport to a research career}}.  The
Ph.D. period often influence a research scholar to make  or to
break in a scientific career. Here one reaches the frontier of 
knowledge and begins in earnest the lifelong task of learning 
how to do research.  As pointed out by Beasley and Jones 
[\ref{beas}] during Ph.D. course ideally one learns how to pick 
a research problem, how to carry out it, how to extract new 
information from the results and how to publish the findings 
to the scientific community.  Thesis or Ph.D. research 
inherently involves those aspects of subject that cannot 
be actually learned  from textbooks or from lecture courses.  
It is the point where the values, traditions and styles of 
science are transmitted from one generation to another.
\subsection{\large{\bf{\emph{Importance of Research}}}}
Research is important both in scientific and nonscientific fields.
In our life new problems, events, phenomena and processes occur
every day.  Practically, implementable solutions and suggestions
are required for tackling new problems that arise.  Scientists
have to undertake research on them and find their causes,
solutions, explanations and applications. Precisely, research
assists us to understand nature and natural phenomena.
 \vskip 10pt
Some important avenues of research are:
\renewcommand{\labelenumi}{(\theenumi)}
\renewcommand{\theenumi}{\arabic{enumi}}
\begin{enumerate}
\item
A research problem refers to a difficulty which a researcher or a
scientific community or an industry or a government organization
or a society experiences.  It may be a theoretical or a practical
situation.  It calls for a thorough understanding and possible
solution.
 \item
Research on existing theories and concepts help us identify the
range and applications of them.
 \item
It is the fountain of knowledge  and provide guidelines for
solving problems.
 \item
Research provides  basis for many government policies. For
example, research on the needs and desires of the people and on
the availability of revenues to meet the needs helps a government
to prepare a budget.
\item
It is important in industry and business for higher gain and
productivity and to improve the quality of products.
 \item
Mathematical and logical research on business and industry
optimizes the problems in them.
\item
It leads to the identification and characterization of new
materials, new living things, new stars, etc.
\item
Only through research inventions can be made; for example, new and
novel phenomena and processes such as superconductivity and
cloning have been discovered only through research.
\item
Social research helps find answers to social problems.  They
explain social phenomena and seek solution to  social problems.
 \item
Research leads to a new style of life and makes it delightful and
glorious.
\end{enumerate}

Emphasizing the importance of research Louis Pasteur said:
{\emph{I beseech you to take interest in these sacred domains
called laboratories.  Ask that there be more and that they be
adorned for these are the temples of the future, wealth and
well-being.  It is here that humanity will learn to read progress
and individual harmony in the works of nature, while humanity's
own works are all too often those of barbarism, fanaticism and
destruction}}. (Louis Paster -- article by S.~Mahanti, Dream 2047,
p.29--34 (May 2003)).
 \vskip 10pt
In order to know what it means to do research one may read
scientific autobiographies like Richard Feynmann's  {\emph{Surely you
are joking, Mr.Feynmann!}}, Jim Watson's {\emph{The double helix}} and
{\emph{Science as a way of life -- A biography of C.N.R. Rao}} by Mohan
Sundararajan.
\sectionmark{RESEARCH METHODS AND RESEARCH METHODOLOGY}
\section{RESEARCH METHODS AND RESEARCH METHODOLOGY}
\sectionmark{RESEARCH METHODS AND RESEARCH METHODOLOGY}
{\emph{Is there any difference between research methods and
research methodology?}}
 \vskip 10pt
{\emph{\bf{Research methods}}} are the various procedures,
schemes and algorithms used in research.  All the methods used
by a researcher during a research study are termed as
{\emph{research methods}}.  They are essentially planned,
scientific and value-neutral. They include theoretical procedures,
experimental studies, numerical schemes,  statistical approaches,
etc. Research methods help us collect samples, data and find a
solution to a problem.  Particularly, scientific research methods
call for explanations based on collected facts, measurements and
observations and not on reasoning alone.  They accept only those
explanations which can be verified by experiments.
 \vskip 10pt
{\emph{\bf{Research methodology}}} is a systematic way to solve a
problem.  It is a science of studying how research is to be
carried out. Essentially, {\emph{the procedures by which
researchers go about their work of describing, explaining and
predicting phenomena are called research methodology.}}  It is
also defined as the study of methods by which knowledge is gained.
Its aim is to give the work plan of research.
\subsection{\large{\bf{\emph{Importance of Research Methodology 
in Research Study}}}}
It is necessary for a researcher to design a methodology for the
problem chosen.  One should note that even if the method
considered in two problems are same the methodology may be
different. It is important for the researcher to know not only the
research methods necessary for the research under taken but also
the methodology. For example, a researcher not only needs to know
how to calculate mean, variance and distribution function for  a
set of data, how to find a solution of a physical system described
by mathematical model, how to determine the roots of algebraic
equations and how to apply a particular method but also need to
know 
\renewcommand{\labelenumi}{(\theenumi)}
\renewcommand{\theenumi}{\roman{enumi}}
\begin{enumerate}
\item
which is a suitable method for the chosen problem?, 
\item
what is the order of accuracy of the result of a method?,
\item
what is the efficiency of the method? 
\end{enumerate}
\noindent and so on. Consideration of
these aspects constitute a research methodology.
 \vskip 10pt
To understand the difference between research methods and
methodology let us consider the problem of finding the roots of
the quadratic equation
\begin{equation}
 ax^2 + bx + c = 0.
\label{eq1}
\end{equation}
The formulas often used for calculating the roots of
eq.(\ref{eq1}) are
\begin{eqnarray}
  x_{+} & = & \frac{-b + \sqrt{b^2-4ac}}{2a} \; ,
  \label{eq2} \\
  x_{-} & = & \frac{-b - \sqrt{b^2-4ac}}{2a} \; \cdot
  \label{eq3}
\end{eqnarray}
These formulas are, however, inaccurate when $ \vert b \vert
\approx \sqrt{b^2-4ac}$.  The equivalent formulas are
\begin{eqnarray}
  x_{+} & = &\frac{-2c}{b + \sqrt{b^2-4ac}} \;,
  \label{eq4} \\
  x_{-} & = &\frac{-2c}{b - \sqrt{b^2-4ac}} \;.
  \label{eq5}
\end{eqnarray}
When $\vert b \vert \approx \sqrt{b^2-4ac}$ one must proceed with
caution to avoid loss of precision.  If $b>0$, then $x_+ $ should
be computed with the formula given by eq.(\ref{eq2}) and $x_-$
should be computed with the formula given by eq.(\ref{eq3}).  If
$b<0$ then $x_+$ should be evaluated using eq.(\ref{eq4}) and
$x_-$ should be evaluated using eq.(\ref{eq5}).  Here the two
formulas constitute the method of finding roots of the equation of
the form given by eq.(\ref{eq1}).  If you use the formulas given
by eqs.(\ref{eq4}--\ref{eq5}) instead of the formulas given by
eqs.(\ref{eq2}--\ref{eq3}) (often used and familiar to us) to
compute the roots then you should clearly explain why the formulas
given by eqs.(\ref{eq4}--\ref{eq5}) were chosen and why the other
formulas given by eqs.(\ref{eq2}--\ref{eq3}) were not considered.
This is what we mean by a research methodology. That is, research
methodology tells you which method or formula or algorithm has to
be used out of the various existing methods or formulas  or
algorithms.
 \vskip 10pt
More precisely, research methods help us get a solution to a
problem.  On the other hand, research methodology is concerned
with the explanation of the following:
\renewcommand{\labelenumi}{(\theenumi)}
\renewcommand{\theenumi}{\arabic{enumi}}
\begin{enumerate}
 \item
Why is a particular research study undertaken?
 \item
How did one formulate a research problem?
 \item
What types of data were collected?
 \item
What particular method has been used?
 \item
Why was a particular technique of analysis of data used?
\end{enumerate}
The study of research methods gives training to apply them to a
problem.  The study of research methodology provides us the
necessary training in choosing methods, materials, scientific
tools and training in techniques relevant for the problem chosen.
 \vskip 10pt
% \newpage
 \hrule
 \vskip 5pt
\noindent{\bf{Assignment:}}
\renewcommand{\labelenumi}{(\theenumi)}
\renewcommand{\theenumi}{\arabic{enumi}}
\begin{enumerate}
 \item
List out at least $10$ methods which you have learned in your UG
and PG courses and write their purpose or application.
\item
Distinguish between research methods and research techniques.
\item
Distinguish between research methods and research methodology with
an example of your own choice.
\end{enumerate}
 \vskip 5pt
  \hrule
\section{TYPES OF RESEARCH}
Research is broadly classified  into two main classes:
\renewcommand{\labelenumi}{\theenumi.}
\renewcommand{\theenumi}{\arabic{enumi}}
\begin{enumerate}
 \item
Fundamental or basic research
\item
Applied research
\end{enumerate}
\subsection{\large{\bf{\emph{Basic Research}}}}
Basic research  is an investigation on basic principles and
reasons for occurrence of a particular event or process or
phenomenon. It is also called {\emph{theoretical research}}. Study
or investigation of some natural phenomenon or relating to pure
science are termed as {\emph{basic research}}.  Basic researches
some times may not lead to immediate use or application.  It is
not concerned with solving any practical problems of immediate
interest.  But  it is original or basic in character. It provides
a systematic and deep insight into a problem and facilitates
extraction of scientific and logical explanation and conclusion on
it. It helps build new frontiers of knowledge. The outcomes of
basic research form the basis for many applied research.
Researchers working on applied research have to make use of the
outcomes of basic research and explore the utility of them.
 \vskip 10pt
Research on improving a theory or a method is also referred as
fundamental research.  For example, suppose a theory is applicable
to a system provided the system satisfies certain specific
conditions.  Modifying  the theory to apply it to a general
situation is a basic research.
 \vskip 10pt
Attempts to find answers to the following questions actually form
basic research. 
\renewcommand{\labelenumi}{\theenumi}
\renewcommand{\theenumi}{$\bullet$}
\begin{enumerate}
\item
Why are materials like that?  
\item
What are they?  
\item
How does a crystal melt? 
\item
Why is sound produced when water is heated?
\item
Why do we feel difficult when walking on seashore? 
\item
Why are birds arrange them in `$>$' shape when flying in a group?
\end{enumerate}
 \vskip 10pt
Fundamental research leads to a new theory or a new property of
matter or even the existence of a new matter, the knowledge of
which has not been known or reported earlier.  For example,
fundamental research on
\renewcommand{\labelenumi}{(\theenumi)}
\renewcommand{\theenumi}{\arabic{enumi}}
\begin{enumerate}
\item
astronomy may leads to identification of new planets or stars in
our galaxy,
\item
elementary particles results in identification of new particles,
\item
complex functions may leads to new patterns or new properties
associated with them,
\item
differential equations results in new types of solutions or new
properties of solutions not known so far,
\item
chemical reactions leads to development of new compounds, new
properties of chemicals, mechanism of chemicals reactions, etc.,
 \item
medicinal chemistry leads to an understanding of physiological
action of various chemicals and drugs,
\item
structure, contents and functioning of various parts of human body
helps us identify the basis for certain diseases.
 \end{enumerate}
\subsection{\large{\bf{\emph{Applied Research}}}}
In an {\emph{applied research}} one solves certain problems
employing well known and accepted theories and principles. Most of
the experimental research, case studies and inter-disciplinary
research are essentially applied research. Applied research is
helpful for basic research.  A research, the outcome of which has
immediate application is also termed as {\emph{applied research}.
Such a research is of practical use to current activity.  For
example, research on social problems have immediate use. Applied
research is concerned with actual life research such as research
on increasing efficiency of a machine, increasing gain factor of
production of a material, pollution control, preparing vaccination
for a disease, etc. Obviously, they have immediate potential
applications.

 \vskip 10pt
Some of the differences between basic and applied research are
summarized in table {\ref{tab1}}. Thus, the central aim of applied
research is to find a solution for a practical problem which
warrants solution for immediate use, whereas basic research is
directed towards finding information that has broad base of
applications and thus add new information to the already existing
scientific knowledge.

\begin{table}[!h]
\caption{Differences between basic and applied researches.}
 \vskip 10pt
\begin{tabular}{lll}
%{\linewidth}{%
%        >{\setlength{\hsize}{1.0\hsize}}X|
%        >{\setlength{\hsize}{1.0\hsize}}X }
\hline
%&  \\
{\emph{Basic research}} & & {\emph{Applied research}} \\
%& & \\
\hline
%& & \\
Seeks generalization & & Studies individual or specific cases
without \\
 & & the objective to generalize \\
%& & \\
Aims at basic processes & & Aims at any variable which makes
the \\
& & desired difference \\
% & & \\
Attempts to explain why  things   happen & & Tries to say how
things can be changed \\
% & & \\
Tries to get all the facts & & Tries to correct the facts which are
problematic \\
% & & \\
Reports in technical language of the topic & & Reports in common
language \\
% & & \\
  \hline
\end{tabular}
\label{tab1}
\end{table}
%
% \vskip 10pt

%
\subsection{\large{\bf{\emph{Normal and Revolutionary Researches}}}}
Basic and applied researches are generally of two kinds:
{\emph{normal research}} and {\emph{revolutionary research}}. In
any particular field, normal research is performed in accordance
with a set of rules, concepts and procedures called a
{\emph{paradigm}}, which is well accepted by the scientists
working in that field.  Normal research is something like
puzzle-solving: interesting, even beautiful, solutions are found
but the rules are remain same.  In this normal research  sometimes
unexpected novel results and discoveries are realized which are
inconsistent with the existing paradigm.  Among the scientist, a
tense situation then ensues, which increases in intensity until a
scientific revolution is reached.  This is marked by a
{\emph{paradigm shift}} and a new paradigm emerges under which
normal scientific activity can be resumed.
\subsection{\large{\bf{\emph{Quantitative and Qualitative Methods}}}}
The basic and applied researches can be {\emph{quantitative}} or
{\emph{qualitative} or even both.  Quantitative research is based
on the measurement of quantity or amount.  Here a process is
expressed or described in terms of one or more quantities.
The result of this research is essentially a number or a set of
numbers. Some of the characteristics of qualitative research/method are:
 \vskip 5pt
\renewcommand{\labelenumi}{\theenumi}
\renewcommand{\theenumi}{$\bullet$}
\begin{enumerate}
 \item
It is numerical, non-descriptive, applies statistics or mathematics and uses numbers.
\item
It is an iterative process whereby evidence is evaluated.
\item
The results are often presented in tables and graphs.
\item
It is conclusive. 
\item
It investigates the {\emph{what}}, {\emph{where}} and {\emph{when}} of decision making.
\end{enumerate}
 \vskip 10pt
Statistics is the most widely used branch of mathematics in quantitative research.  It finds applications not only in physical sciences but also in economics, social sciences and biology.  Quantitative research using statistical methods often begins with the  collection of data based on a theory or hypothesis or experiment followed by the application of descriptive or inferential statistical methods.
 \vskip 10pt
Qualitative research is concerned with qualitative phenomenon involving quality.  Some of the characteristics of qualitative research/method are:
 \vskip 5pt
\renewcommand{\labelenumi}{\theenumi}
\renewcommand{\theenumi}{$\bullet$}
\begin{enumerate}
 \item
It is non-numerical, descriptive, applies reasoning and uses words.
\item
Its aim is to get the meaning, feeling and describe the situation.
\item
Qualitative data cannot be graphed.
\item
It is exploratory. 
\item
It investigates the {\emph{why}} and {\emph{how}} of decision making.
\end{enumerate}
\vskip 10pt
We measure and weigh things in the
study of substance or structure.  Can we measure or weigh
patterns?  We cannot measure or weigh patterns.   But to study
patterns we must map a configuration of relationships.  That is,
structures involve quantities whereas patterns involve qualities.
If one wishes to investigate why certain data are random then it
is a qualitative research.  If the aim is to study how random the
data is, what is the mean, variance and distribution function then
it becomes quantitative.  Explaining how digestion of food takes
place in our body is a qualitative description.  It does not
involve any numbers or data and quantities.
 \vskip 10pt
The detection of a particular compound is a qualitative analysis.
This can be done by carrying out physical or chemical tests.
Determination of exact amount of a particular compound present in
a volume is essentially quantitative analysis. This can be done by
volumetric, gravimetric and colorimetric methods or instrumental
methods. Experimental and simulation studies are generally
quantitative research.
 \vskip 10pt
In fact, qualitative methods can be used to understand the meaning of the numbers obtained by quantitative methods.
\subsection{\large{\bf{\emph{Other Types of Research}}}}
Other types of research include {\emph{action research}} (fact
findings to improve the quality of action in the social world),
{\emph{explanatory research}} (searching explanations for events
and phenomena, for example finding answer to the question why are
the things like what they are?), {\emph{exploratory research}}
(getting more information on a topic) and {\emph{comparative
research}} (obtaining similarities and differences between events,
methods, techniques, etc.).  For discussion on these types of
research see refs.[\ref{kothari}--\ref{phil}].
%
%\newpage
 \vskip 10pt
 \hrule
 \vskip 5pt
\noindent{\bf{Assignment:}}
\renewcommand{\labelenumi}{(\theenumi)}
\renewcommand{\theenumi}{\arabic{enumi}}
\begin{enumerate}
\addtocounter{enumi}{3}
 \item
List out at least $10$ theoretical and applied methods which you
have learned in your UG, PG courses and write their features in
two or three sentences.
 \item
Write at least $20$ questions in your subject the investigation of
which forms basic research.  Then point out how many of them have
already been solved and how many were found  in applications.
 \item
Distinguish between theory and experiment.
 \item
Write a note on importance of theory in basic and applied
researches.
 \item
Bring out the importance of inter-disciplinary research.
\end{enumerate}
 \vskip 5pt
  \hrule
\section{ENTERING INTO RESERCH}
\noindent {\emph{How do you enter into a research career?}}
 \vskip 5pt
There are many ways to enter and start a research career.  In India, one popular path is to appear for the National Eligible Test (NET) conducted by the National Education Testing Bureau of the University Grants Commission (UGC).  This test is conducted twice in a year generally in June and December.  The NET is conducted in humanities, languages, social sciences, forensic science, environmental sciences, computer science and applications and electronics.  The Council of Scientific and Industrial Research (CSIR) conducts the UGC--CSIR NET for science subjects like mathematical, physical, chemical, life, earth, atmospheric, ocean and planetary sciences--jointly with the UGC.  
 \vskip 10pt
One of the prime objectives of the NET is to ensure minimum standards for the entrants in the research.  Those who have at least 55 percent of marks in their postgraduate degree are eligible for writing the test.  Those who are appearing for the final-year qualifying examination can also apply for the test under the Result Awaited category.  Age limit for JRF is 28 years.  The upper age limit may be relaxed up to five years for SC/ST/OBC/PH and female applicants.  For more details, visit www.csirhrdg.res.in.
\vskip 10pt
Passing the test means one is eligible for the award of Juniour Research Fellowship (JRF) for a period of five years in a university or a research institution or a college.
 \vskip 10pt
Research facilities are availble in research institutions and CSIR laboratories for those who got good grades in the Graduate Aptitude Test in Engineering (GATE) conducted by the Indian Institutes of Technology (IITs).  There is another possible path to enter research.  Scientists working in research and educational institutes prepare research proposal and submit to government agencies like Department of Science and Technology (DST), CSIR, UGC, Department of Atomic Energy (DAE), National Board for Higher Mathematics (NBHM), etc.  Generally, JRF and other higher fellowships are proposed by the proposer to carry out the proposed research work.  Once the proposal is approved then advertisement will be given in newspapers to apply for the research fellowships.  Many universities also provide limited number of fellowships.  In the above routes a researcher will get fellowship to do research.  Without fellowship also one can start a research career.  However, since research period for Ph.D. degree is generally a 4--6 years of work, it is not advisable to start a research life without a fellowship.
\section{VARIOUS STAGES OF A RESEARCH}
Whenever a scientific problem is to be solved there are several
important steps to follow.  The problem must be stated clearly,
including any simplifying assumptions.  Then develop a
mathematical statement of the problem.  This process may involve
use of one or more mathematical procedures.  Frequently, more
advanced text books or review articles will be needed to learn
about the techniques and procedures.  Next, the results have to be
interpreted to arrive at a decision.  This will require experience
and an understanding of the situation in which the problem is
embedded. A general set of sequential components of research is
the following:
\renewcommand{\labelenumi}{\theenumi.}
\renewcommand{\theenumi}{\arabic{enumi}}
\begin{enumerate}
 \item
Selection of a research topic
 \item
Definition of a research problem
 \item
Literature survey and reference collection
 \item
Assessment of current status of the topic chosen
 \item
Formulation of hypotheses
 \item
Research design
 \item
Actual investigation
 \item
Data analysis
 \item
Interpretation of result
 \item
Report
 \end{enumerate}
In the following sections the above mentioned various stages of
research are discussed in detail.
\sectionmark{SELECTION OF A RESEARCH TOPIC AND PROBLEM}
\section{SELECTION OF A RESEARCH TOPIC AND PROBLEM}
\sectionmark{SELECTION OF A RESEARCH TOPIC AND PROBLEM}
The starting point of a research is the selection of a research
topic and problem.  History teaches the continuity of the development and progress of science.  The point is that every age has its own problems, numerous in number, which the following age either solves or casts aside as profitless and replaces by new one.  If we could obtain an idea of the probable development of scientific knowledge in the immediate future, we must let the unsettled questions pass before our minds and look over the problems which the science of today sets and whose solution we expect from the the near future.  The deep significance of certain problems for the advancement of science and society must be taken into consideration in choosing a problem of research.

 \vskip 10pt
There are many ways to do research as there 
are scientists.  The choice of a thesis research area and adviser 
has always been a very subjective process. Identifying a suitable 
topic for work is one
of the most difficult parts of a research. Before choosing a
research topic and a problem the young researchers should keep the
following  points in mind.
 \renewcommand{\labelenumi}{\theenumi}
\renewcommand{\theenumi}{$\bullet$}
\begin{enumerate}
 \item
Topic should be suitable for research.
 \item
The researcher should have interest in it.
 \item
Topic should not be chosen by compulsion from some one else.
 \end{enumerate}
Topic and problem can be fixed in consultation with the research
supervisor.  In our country often research supervisors suggest a
topic and state a problem in broad view.  The researcher has to
narrow it and define it in an operational form. One may ask: Is it
necessary that the topic of a Ph.D.  should be different from
M.Sc. project and M.Phil dissertation?  The answer is not
necessary.  If a student is able to get a supervisor working in
his M.Sc.project or M.Phil dissertation topic then it would save
about six months in the duration of his Ph.D. work.
\subsection{\large{\bf{\emph{Can a Researcher Choose a Topic 
 by himself?}}}}
 \label{s1}
A youngster interested to start a research career wishes to know
whether he/she has freedom to do research in the topic of his/her
own interest. The style of research in our country and various
other factors like the infrastructure facility available in a
research institute, time limit, our commitment to family and
social set up hardly allow a young researcher to choose a topic by
himself for his PG project, M.Phil. dissertation and Ph.D. thesis.
However, many research supervisors give complete freedom to choose
a problem in the topic suggested by him for a Ph.D. research work.
Because the normal time duration of M.Phil dissertation is about
6-8 months, it is better to work on the problem suggested by the
supervisor.
 \vskip 10pt
If a student wishes to do research (for Ph.D. degree) with
fellowship then he cannot have freedom to choose a topic since he
has to work on a project the goal of which is already defined by
the project investigator.  In the other way, after choosing a
topic of his own interest he has to find a supervisor who is
working in that topic or interested in guiding him.  In this case
one has severe limitation in our country for getting a fellowship
and for registering for a research degree. If a student is not
very much particular about the fellowship he has a chance to do
research in the topic of his own interest. A researcher in India
after two years of research experience with few (two or more)
publications can apply for a senior research fellowship (SRF) to
CSIR (for details
see its and other relevant web sites).  He can prepare a project
under the direction of his Ph.D. supervisor which can lead to a
fellowship.  For details see the book {\emph{How to Get Scholarships,
Fellows and Stipends}} by  K.D.~Kalaskar (Sultan Chand and Sons, New
Delhi).
 \vskip 10pt
Considering the above, a researcher should make-up his mind so as
to work in a topic suggested by the supervisor.  However, a
research problem may be chosen by a researcher himself. This has
several advantages.  In this case
\renewcommand{\labelenumi}{\theenumi}
\renewcommand{\theenumi}{$\bullet$}
\begin{enumerate}
 \item
the researcher can pursue his/her own interest to the farthest
limits,
 \item
there is an opportunity to spend a long time on something that is
a continuous source of his pleasure and
 \item
the results would prove better in terms of the growth of the
investigator and the quality of the work.
 \end{enumerate}
If the researcher is not interested in the topic and problem
assigned to him but he is working on it because of supervisor's
compulsion, then he will not be able to face and overcome the
obstacles which come at every stage in research.
\subsection{\large{\bf{\emph{Identification of a Research Topic 
and Problems}}}}
Some sources of identification of a research topic and problems
are the following:
\renewcommand{\labelenumi}{(\theenumi)}
\renewcommand{\theenumi}{\arabic{enumi}}
\begin{enumerate}
 \item
Theory of one's own interest
 \item
Daily problems
 \item
Technological changes
 \item
Recent trends
 \item
Unexplored areas
 \item
Discussion with experts and research supervisor
\end{enumerate}
Suppose one is interested in the theory of nonlinear differential
equations or quasicrystals or fullerenes.  Then he can find a
research guide who is working in this field or interested to work
in this field and then choose a problem for research.
 \vskip 10pt
Our daily experiences and day to affairs have rich openings on
various aspects such as the daunting tasks of AIDS, air pollution,
afforestation and deforestation, child labor, problems of aged
citizens, etc.
 \vskip 10pt
Technology in various branches of science, business and marketing
changes rapidly.  For example, in the early years, computers were
built in larger size with vacuum tubes.  Then evolution in
electronic technology replaced them by integrated circuits.
Recently, scientists have developed quantum dots.  Now the
interest is in developing efficient, super-fast and miniaturized
computing machine made up of material whose particle size of the
order of nano ($10^{-9}$) meter or even smaller.  Similarly,
another fascinating topic namely, {\emph{thin film}} has multiple
fields of applications.  Recent research on fullerenes resulted in
many practical applications.
 \vskip 10pt
Choosing a topic of current interest or recent trends provides
bright and promising opportunities for young researchers to get
post-doctoral fellowship, position in leading institutions in our
nation and abroad.
 \vskip10 pt
In each subject there are several topics which are not explored in
detail even though the topic was considered by scientists long
time ago. For example, string theory, quantum computing,
nano-particles, quantum cloning and quantum cryptography and gene
immunology are fascinating topics and are in preliminary stages.
 \vskip 10pt
The supervisors and experts are working on one or few fields over
a long time and they are the specialists in the field considered
and well versed with the development and current status of the
field. Therefore, a young researcher can make use of their
expertise in knowing various possible problems in the topic the
solving of which provide better opportunities in all aspects.
 \vskip 10pt
Don't choose a topic simply because it is fascinating.  In
choosing a topic one should take care of the possibility of data
collection, quantity of gain, breadth of the topic and so on.  The
topic should not be too narrow.  For example, the study of social
status and sexual life of married couples of same sex (man-man
marriage and woman-woman marriage) is interesting and of social
relevance.  But the intricate problem here is that we do not find
enough number of such couples to study.  This is a very narrow
topic at the same time we will not get enough data to analyze. On
the other hand, the changes in the social life of aravanis in
recent times is a valuable social problem and one can collect
enough data.
 \vskip 10pt
Further, one has to study advanced level text books and latest
research articles to identify  problems. Is it necessary to know
all the methods, techniques, concepts in a research topic before
identifying a problem for investigation?  This is not necessary.
After learning some fundamental concepts, recent developments and
current trends of a topic, one can identify a problem for
research. Then he can learn the tools necessary to solve it.
\subsection{\large{\bf{\emph{Definition and Formulation of a
Problem}}}}}
After identifying a problem, in order to solve it, it has to be
defined and formulated properly.  For this purpose, one can
execute the following.
\renewcommand{\labelenumi}{\theenumi}
\renewcommand{\theenumi}{$\bullet$}
\begin{enumerate}
 \item
State the problem in questionnaire form or in an equivalent form
 \item
Specify the problem in detail and in precise terms
 \item
List the assumptions made
 \item
Remove the ambiguities, if any, in the statement of the problem
 \item
Examine the feasibility of a particular solution
 \end{enumerate}
\noindent{Defining the problem is more important than its
solution.  It is a crucial part of the research study and should
not be defined in hurry.}

\subsection{\large{\bf{\emph{How do you Asses Whether the Defined
Problem as a Good Problem?}}}}
A problem in its first definition may not be appealing.  It may
require redefinition in order to make it a good problem.  That is,
by suitably rewording or reformulating the chosen problem, it can
be made to meet the criteria of a good problem.  This is also
important to solve the problem successfully.  To this end a
researcher can ask a series of questions on the problem. Some are:
\renewcommand{\labelenumi}{(\theenumi)}
\renewcommand{\theenumi}{\arabic{enumi}}
\begin{enumerate}
 \item
Is the problem really interesting  to him and to the scientific
community?
 \item
Is the problem significant to the present status of the topic?
 \item
Is there sufficient supervision/guidance?
 \item
Can the problem be solved in the required time frame?
 \item
Are the necessary equipments, adequate library and computational
facilities, etc. available?
\end{enumerate}
If the answers to these questions are satisfactory, then the
researcher can initiate work on the chosen problem.  In addition,
discuss the problem with the current doctoral students and obtain
the scope of the problem and other related aspects.
\subsection{\large{\bf{\emph{How are these Questions Important 
and Relevant to a Researcher?}}}}
The researcher should be interested on the problem for the reasons
mentioned earlier at the end of the Sec.(\ref{s1}).  The problem
should also be interesting to the supervisor so that the
researcher can get the necessary guidance from him.  Otherwise
sometimes the researcher may find it very difficult to convince
the supervisor on the importance and significance of the results
obtained.  More importantly, the problem must be of interest to
scientific community and society. If not then the researcher will
find great difficulty to publish his findings in reputed journals
and convince the funding agency.
 \vskip 10pt
Next, the status of the problem, particularly the importance of
finding its solution should match with the current status of the
field.   But, if the problem investigated is of not much interest
to science and society then publications will become useless to
him in his research career.  Specifically, they cannot help earn a
post-doctoral fellowship, respectability and a permanent job in an
institution.
 \vskip 10pt
A researcher needs proper guidance and encouragement from the
supervisor regularly.  This is important for keeping the research
in right track, to overcome the difficulties which come at various
states of research and also to have moral support.  A researcher
should avoid working under the guidance of a supervisor having
serious health problems or family problems, committed his large
time to administrative work and strong involvement in nonacademic
matters.
 \vskip 10pt
The following story was told by S.L.~Glashow (Harvard University)
[Julian Schwinger: The Physicist, the Teacher, and the Man. (Ed.)
Y.~Jack Ng, World Scientific, Singapore, 1996, pp.155]:

Once upon a time, a fox came upon a rabbit who was typing away in
the middle of the forest.  {\emph{What do you think you are
doing?}} asked the fox.  {\emph{I am writing my thesis on how
rabbits eat foxes}} replied the rabbit.  {\emph{Nonsense}}! said
the fox, {\emph{rabbits don't eat foxes; foxes eat rabbits}}.
{\emph{Just take a peek in my cave}} challenged the rabbit. The
fox entered the rabbit's cave and was never seen again.  Some
time, later a wolf came to the rabbit, who was still writing his
thesis.  {\emph{What do you thing you are doing?}} said the wolf,
and a similar interchange took place.  The wolf entered the cave
and was never seen again.  Finally, a bear came to chat with the
rabbit.  {\emph{I am writing my thesis on how rabbits eat bears}}
said the rabbit. {\emph{Nonsense}}! growled the bear, who
was sent to the cave never to be seen again.  A wise owl watched
these strange goings-on and was puzzled.  Softly sneaking into the
rabbits cave, he came upon a neat pile of fox bones.  A bit
further on, he discovered a neat pile of wolf bones.  Finally, at
the back of the cave behind a neat pile of bear bones, sat a very
fat and satisfied lion picking his teeth with a bear claw.  The
moral of the story is that {\emph{it really doesn't matter what
your thesis subject is.  What counts is your choice of an
advisor}}.
 \vskip 10pt
An important point is that before initiating research work on
a problem, a rough estimate on costs and time required to complete
the work must be made.  A problem suitable for Ph.D. degree should
not be taken for M.Phil. degree.  A problem suitable for M.Phil.
degree is not appropriate for Master's degree. If the collection
of data or resources or related information takes many years, then
the topic is obviously inappropriate for Ph.D. degree.
Controversial subjects should not be chosen. Problems that are too
narrow or too vague should be avoided.
 \vskip 10pt
Finally, the researcher must make sure that the necessary
experimental setup and materials to perform the actual research
work are available in the department where research work is to be
carried out.  Without these, if the researcher initiated the work
and has gone through certain stages of work or spent one or two
years in the problem then in order to complete the task he would
be forced to buy the materials and instruments from his personal
savings.
\section{LITERATURE SURVEY}
After defining a problem, the researcher has to do literature
survey connected with the problem. {\emph{Literature survey is a
collection of research publications, books and other documents
related to the defined problem}}.  It is very essential to know
whether the defined problem has already been solved, status of the
problem, techniques that are useful to investigate the problem and
other related details. One can survey
\renewcommand{\labelenumi}{(\theenumi)}
\renewcommand{\theenumi}{\arabic{enumi}}
\begin{enumerate}
 \item
the journals which publish abstracts of papers published in
various journals,
 \item
review articles related to the topic chosen,
 \item
journals which publish research articles,
  \item
advanced level books on the chosen topic,
 \item
proceedings of conferences, workshops, etc.,
 \item
reprint/preprint collections available with the supervisor and
nearby experts working on the topic chosen and
 \item
Internet.
\end{enumerate}
 \vskip 5pt
A free e-print service provider for physics, mathematics,
nonlinear science, computer science and biology is
 \vskip 3pt
http://www.arXiv.org
 \vskip 5pt
No research shall be complete unless we make use of the knowledge
available in books, journals and internet.  Review of the
literature in the area of research is a preliminary step before
attempting to plan the study.
 \vskip 5pt
Literature survey helps us
\renewcommand{\labelenumi}{(\theenumi)}
\renewcommand{\theenumi}{\arabic{enumi}}
\begin{enumerate}
\item
sharpen the problem, reformulate it or even leads to defining
other closely related problems,
\item
get proper understanding of the problem chosen,
\item
acquire proper theoretical and practical knowledge to investigate
the problem,
\item
show how the problem under study relates to the previous research
studies and
\item
know whether the proposed problem had already been solved.
\end{enumerate}
Through survey one can collect relevant information about the
problem. Clarity of ideas can be acquired through study of
literature.
 \vskip 5pt
Apart from literature directly connected with the problem, the
literature that is connected with similar problems is also useful.
It helps formulate the problem in a clear-cut way. A review on
past work helps us know the outcome of those investigations where
similar problems were solved.  It can help us design methodology
for the present work.  We can also explore the vital links with
the various trends and phases in the chosen topic and familiarize
with characteristic precepts, concepts and interpretations.
Further, it can help us formulate a satisfactory structure of the
research proposal.
 \vskip 10pt
Because a Ph.D. thesis or M.Phil. dissertation is a study in depth
aiming contribution to knowledge, a careful check should be made
to ensure that the proposed study has not previously been
performed and reported. The earlier studies which are relevant to
the problem chosen should be carefully studied. Ignorance of prior
studies may lead to a researcher duplicating a work already
carried out by another researcher.  A good library will be of
great help to a researcher at this stage. One can visit nearby
research institutions and avail the library facility. Review the
latest research papers and Ph.D. theses to acquire recent trends.
\section{THE INTERNET AS A MEDIUM FOR RESEARCH}
From past one decade or so the internet became an important source of knowledge and an effective medium for research.  For researchers, it is providing a range of new opportunities for collecting information, networking, conducting research, collecting data and disseminating research results. 
 \vskip 10pt
Electronic mail, e-journal, on-line submission of articles to journals, online focus groups, online video conferencing and online questionary are some of the latest tools opened-up by the internet.  
We note that thousands of books and other print publications have been made  available online that would be extremely difficult to locate otherwise, including out-of-print books, and classic literature and textbooks that would be much less accessible in their printed form.
 \vskip 10pt
Some of the scientific research information available on the internet are:
\renewcommand{\labelenumi}{\theenumi}
\renewcommand{\theenumi}{$\bullet$}
\begin{enumerate}
 \item
Details about various scientific and nonscientific topics.
\item
Titles and other relevant information of article published in various journals, possibly, from past one decade or so (full article will not be available).
\item
Preprint of papers submitted by researchers in certain websites.
\item
Information about scientific meetings to be held.
\item
Contact details for other researchers.
\item
Databases of reference material.
\item
Places where one can discuss topics and ask for help.
\end{enumerate}
\noindent In general, academic research that has been commercially published is not freely available on the internet.
 \vskip 10pt
Some of the features of conducting research through internet are:
\renewcommand{\labelenumi}{\theenumi}
\renewcommand{\theenumi}{$\bullet$}
\begin{enumerate}
 \item
Short time for collecting and recording data.
\item
Data unknown to us can be identified and downloaded.
\item
The possibility of conducting interviews and focus groups by e-mail, which results in enormous saving in costs and time.
\item
New communities to act as the object of social scientific enquiry.
\end{enumerate}
\vskip 10pt
While the internet contains a virtually-unlimited wealth of information not found in traditional resources, this abundance also may hinder academic research.  The following are some of the new challenges for the researcher:
\renewcommand{\labelenumi}{\theenumi}
\renewcommand{\theenumi}{$\bullet$}
\begin{enumerate}
 \item
Problems of sampling.
\item
Reliability and accuracy of the obtained data information.
\item
The ethics of conducting research into online communities.
\item
Physical access and skills required to use the technologies involved.
\item
The changed chronology of interaction resulting from asynchronous communication.
\end{enumerate}
 \vskip 5pt
A major way to find whether an online source is credible is to determine how popular and authoritative the source is.  If the site has a well-respected offline counterpart such as the New York Times the site will be as credible as the original.  For websites and authors which have little popularity, one must consider the credentials of the source--if those are available and valid.  Even though a website may be written in a professional or academic manner, the lack of central body to determine its credibility may be a prohibitive factor for serious research.
 \vskip 10pt
An example of an online research in which researchers have used the internet as a medium for performing research is {\emph{National 
Geographic Survey 2000}}.  In this survey, interactive survey instruments were used to study and analyse the effects of location and mobility on values and cultural tastes.  Extensive use was made of public relations and community outreach to publicize the survey.  In about two months time 80,000 self-selected participants from 178 countries started the questionnaire and 55,000 of them completed it.  Questions were in some measure dynamically generated, with respondents automatically redirected to appropriate section based on their answers.  The main survey site is
 \vskip 5pt
 http://survey2000.nationalgeographic.com/
\vskip 10pt
Some of the potential advantages of online questionnaire are:
\renewcommand{\labelenumi}{\theenumi}
\renewcommand{\theenumi}{$\bullet$}
\begin{enumerate}
 \item
Low-cost delivery and return.
\item
Wide potential coverage.
\item
Ease of completion.
\item
Submission and data capture.
\item
Appropriateness to particular populations.
\item
high respondent acceptance for some groups.
\end{enumerate}
 \vskip 10pt
Potential difficulties include:
\renewcommand{\labelenumi}{\theenumi}
\renewcommand{\theenumi}{$\bullet$}
\begin{enumerate}
 \item
The paucity of methodological literature.
\item
Appropriateness to research aims.
\item
Target population.
\item
Technical difficulties.
\item
Sampling and response rates.
\end{enumerate}
\section{REFERENCE COLLECTION}
As soon as the survey of available source begins, the preparation
and collection of references preferably with annotations should be
undertaken.  Keeping records systematically during research helps a researcher achieve various objectives. It preserves data for future use.
The researcher may stumble upon something that may not be of immediate use, but would help him later.  Details have to be kept in files.
 \vskip 10pt
The important source of reference collection is the
journal called {\emph{Current Contents}}.  This comes once in a week.  It is available in hard copy and also in floppy diskette. Almost all
the universities and research institutions buy this document. It contains
the table of content of research journals and magazines in various
subjects. It provides title of articles, names of the authors,
date of publication, volume number, starting page number of the
articles and address of the author from whom one can get the
reprint of the article.  If the title of the article indicates
that the paper is in the topic of one's interest then he can take
a copy of the article if the journal is available in the local
library. Otherwise, he can get it from a document delivery service
centre. For example, in India INFLIBNET provides this service
through six institutions.  For details visit the following web
sites:
 \vskip 3pt
http://web.inflibnet.ac.in/index.isp
 \vskip 3pt
http://www.iisc.ernet.in/
 \vskip 3pt
http://www.jnu.ac.in/
 \vskip 3pt
\noindent One can obtain a research article on paying the charge
fixed by the INFLIBNET provided the particular journal is
available in it. Articles can also be purchased from the
publishers on payment. Alternatively, reprint of the article can
be had from the author by sending a letter/card/e-mail to the author.
 \vskip 10pt
The references from current contents or from journals can be noted
on a separate card or sheet with the names of authors and the
title of the paper/book, etc.  For a research paper, its title,
journal name, volume number, starting and ending pages of it and
year of publication should be noted. For a book, publisher's name,
place of publication and year of publication must be written down.
Instead of cards, nowadays one can store the details of the
references in computers and have a copy in two or three floppy
diskette.  The references can be classified.  For example, sources
dealing with theory, dealing with experimental techniques,
concerned with numerical methods, etc. can be grouped separately.
The copies of the research articles can also be classified and
bounded.  Cross references (that is research articles or books
referred or cited in a research report) should also be collected
and classified. These also provide useful information.
 \vskip 10pt
Reference collection and keeping the collected materials have to be systematic.  Unless they are organized with utmost care and discipline, one would end up in chaos. One may not be able to retrieve the required research article or any other collected material when it needed.   Materials can be classified as facts, ideas, views and opinions, expert comments, new breakthroughs, quotes, journal papers, review articles, etc.  It is better to have multiple copies of important materials.    At various stages of research one may refer to numerous journal articles, books and web sites.  Obviously, all of them are not going to find a place in the thesis or research reports.  Based on the present work and future plan one has to select the relevant materials from the available collection.
\section{ASSESSING THE CURRENT STATUS}}
Generally, it is not difficult to know the current status of
research work in a specific topic.  The current status of the
chosen topic can be identified by reading the relevant journals
and the recent papers, discussions in conferences, seminars and
workshops.  One can perform inquiries at several important places
known for research on proposed topic.
 \vskip 10pt
A study of the current literature in the chosen topic explores the
current status of it.  More importantly,  review articles point
out not only to the basic aspects and features of the topic
concerned but also give a brief account of its present status. For
this purpose, one can survey the journals (for a topic in physics)
such as Physics Reports, Reviews of Modern Physics, Physical
Review Letters, Review section of American Journal of Physics,
Pramana, Current Science and  Proceedings of recently conducted
seminars and conferences, etc.
 \vskip 10pt
Rapid communication and Letter sections of international journals
publish articles which are very important and fall in recent
trends category.  There are several areas in internet where the
papers just submitted to journals are placed.  One can download
such articles free of cost.  These articles indicate the recent
trends in a particular topic.  Some relevant web sites are listed
below.
\vskip 10pt
 http://arxiv.org/
 \vskip 3pt
 http://www.ams.org/global-preprints/
 \vskip 3pt
 http://front.math.ucdavis.edu/math.AG/
 \vskip 3pt
 http://www.ma.utexas.edu/m$\mathrm{p}_{-}$arc/
 \vskip 3pt
 http://www.clifford.org/anonftp/clf-alg/

\section{HYPOTHESIS}
Researchers do not carry out work without any aim or expectation.
Research is not of doing something and presenting what is done.
Every research problem is undertaken aiming at certain outcomes.
That is, before starting actual work such as performing an
experiment or theoretical calculation or numerical analysis, we
expect certain outcomes from the study.  The expectations form the
hypothesis.  {\emph{Hypotheses are scientifically reasonable
predictions}}. They are often stated in terms of if-then sentences
in certain logical forms.  A hypothesis should provide what we
expect to find in the chosen research problem.  That is,
the expected or proposed solutions based on available data and
tentative explanations constitute the hypothesis.
 \vskip 10pt
Hypothesizing is done only after survey of relevant literature and
learning the present status of the field of research. It can be
formulated based on previous research and observation. To
formulate a hypothesis the researcher should acquire enough
knowledge in the topic of research and a reasonably deep insight
about the problem. In formulating a hypothesis construct
operational definitions of variables in the research problem.
Hypothesis is due to an intelligent guess or for inspiration which
is to be tested in the research work rigorously through
appropriate methodology.  Testing of hypothesis leads to
explanation of the associated phenomenon or event.
 \vskip 10pt
{\emph{What are the criteria of a good hypothesis?}}   An
hypothesis should have conceptual clarity and a theoretical
orientation. Further, it should be testable.  It should be stated
in a suitable way so that it can be tested by investigation.  A
hypothesis made initially may become incorrect when the data
obtained are analyzed.  In this case it has to be revised.  It is
important to state the hypothesis of a research problem in a
research report. We note that if a hypothesis withstands the
experiments and provides the required facts to make it acceptable,
not only to the researchers performing the experiments but to
others doing other experiments then when sufficiently reinforced
by continual verification the hypothesis may become a
{\emph{theory}} [\ref{span}].
 \vskip 10pt
According to Poincar\'{e}, {\emph{a scientific hypothesis which
was proved untenable can still be very useful.  If a hypothesis
does not pass an empirical test, then this fact means that we have
neglected some important and meaningful element.  Thus, the
hypothesis gives us the opportunity to discover the existence of
an unforeseen aspect of reality}}.  As a consequence of this point
of view about the nature of scientific theories,  Poincar\'{e}
suggested that a scientist must utilize few hypotheses, for it is
very difficult to find the wrong hypothesis in a theory which
makes use of many hypotheses.
\section{MODE OF APPROACH}
Mode of approach means the manner in which research is to be
carried out. {\emph{It should keep the researcher on the right
track and make him complete the planned work successfully}}. 
You should keep in mind that there is always room for improvement in any human endeavor, and research is no exception.  Do each and every job with maximum care.  If you go for endless fine-tuning, you will never finish the job on time.  You have to be pragmatic in your approach and execution.  One
should sharpen the thinking and focus attention on the more
important aspects of the study. The scientific thinking must be
more formal, strict, empirical and specific and more over goal
oriented. Essentially, one must concentrate on an area of research
and aim to perform better than almost anyone else. In order to
make steady progress in research and to asses the progress of the
research work, a research design is very helpful.
\subsection{\large{\bf{\emph{Research Design}}}}
{\emph{Plan your work and work your plan}} is the suggestion of
Napolean Hill. For a scientific research one has to prepare a
research design. It should indicate the various approaches to be
used in solving the research problem, sources and information
related to the problem and, time frame and the cost budget.
Essentially, the research design creates the foundation of the
entire research work. The design will help perform the chosen task
easily and in a systematic way. Once the research design is
completed the actual work can be initiated. The first step in the
actual work is to learn the facts pertaining to the problem.
Particularly, theoretical methods, numerical techniques,
experimental techniques and other relevant data and tools
necessary for the present study have to be collected and learnt.
 \vskip 10pt
It is not necessary that every theory, technique and information
in the topic of research is useful for a particular problem. A
researcher has to identify and select materials which are useful
to the present work.  Further, the validity and utility of the
information gathered should be tested before using them.
Scientific research is based on certain mathematical, numerical
and experimental methods.  These sources have to be properly
studied and judged before applying them to the problem of
interest.
\subsection{\large{\bf{\emph{What are the Possible Approaches 
to be Followed by a Researcher?}}}}
Being a member of a research institution alone is  not sufficient to become a scientist.  
Every great human achievement is preceded by extended periods of
dedicated and concentrated effort.  As told by Mahatma Gandhi
{\emph{satisfaction lies in the effort and not in the
attainment}}. Full effort is full victory. A researcher can
exercise the following aspects regularly throughout the research
carrier. These will keep him in right track and tightly bind him
to the research activity.
\renewcommand{\labelenumi}{(\theenumi)}
\renewcommand{\theenumi}{\arabic{enumi}}
\begin{enumerate}
 \item
Discussion with the supervisor, experts and colleagues about the
research work, particularly, the problem and its origin,
objectives and difficulties faced in the execution of the problem.
 \item
Reading of the latest research papers, relevant theories and
possible application to the present problem and to overcome the
difficulties faced.
 \item
Review of the work reported on the similar problems.
 \item
Theoretical calculations, setting-up of an experimental setup,
numerical calculations, computer programs, preparation of graphs,
tables and other relevant work related to the research should be
done by a new researcher by himself without assistance from
others.
 \item
Have a practice of periodically writing the work done, results
obtained and steps followed in a work.  This is important because
sometime we may think that a particular aspect will be a center
piece of the problem under investigation.  But once we make a
write-up of it, this aspect or part of it may turn out to be only
of marginal importance.  In fact, writing of the progress of the
work will help us better understand our work and forms a solid
basis for further progress.  It also points out the gaps in our
work.
 \item
Participation and presentation of research findings in national
and international meetings.
\end{enumerate}

\noindent These regular practices provide useful information like
new ideas and can help the researcher
\renewcommand{\labelenumi}{(\theenumi)}
\renewcommand{\theenumi}{\arabic{enumi}}
\begin{enumerate}
 \item
sharpen and focus attention,
 \item
confining to the formulation and
 \item
in the interpretation of the solution obtained.
\end{enumerate}
 \vskip 10pt
\subsection*{Independent Research}
 \vskip 10pt
Absolute honesty, patience, stamina, precision and devotion to the subject of matter together with imagination and analytical ability are among the requirements for scientific research.
 \vskip 10pt
Each  and every bit of task related to the research work has to be
done by the researcher. A young researcher should not do the
entire work in collaboration with others.  A young researcher should 
have the ability {\emph{to do all by himself}}.  In this connection 
Beasley and Jones [\ref{beas}] wrote: {\emph{In reality collaborations 
are commonplace, often necessary--to get samples, to make all the 
desired measurements or to perform a complete analysis.  But such 
collaborations generally arise naturally in the course of the 
research and define and limit themselves naturally.  A true 
sense of camaraderie often develops, and students do not lose 
that sense of having made major contributions on their own}}.
\vskip 10pt
The researcher is
advised to perform all the works starting from identification of
the problem to report preparation by himself under the guidance of
supervisor. Particularly, collaboration work with experts and
senior researcher may be avoided. (However, he can discuss his
problems with them). This is important to acquire
\renewcommand{\labelenumi}{(\theenumi)}
\renewcommand{\theenumi}{\arabic{enumi}}
\begin{enumerate}
 \item
enough knowledge,
 \item
confidence and
 \item
training
\end{enumerate}
to carry out research independently after getting a Ph.D. degree.
 \vskip 10pt
Part of the dissertation should demonstrate the researcher's
originality.  The dissertation should reflect the efforts of a
single researcher.  Keeping this in mind one should avoid
collaboration as far as possible in the young stage.  For example,
Landau never did for his students what he believed they should do
themselves.  Some times, after many unsuccessful attempts to solve
a problem, a student would ask Landau for his help.  Landau would
reply, {\emph{This is your problem.  Why should I do it for
you?}}  Neither did Landau formulate problems for his students nor
give title of the research work.  In these ways he trained his
students to be {\emph{independent}} and become {\emph{future
leaders of science}}.  Eugene Wigner, a Nobel laureate said:
{\emph{One does not have the satisfaction which creative work provides,
if one's activities are too closely directed by others}}.
 \vskip 10pt
Prof.Balaram wrote {\emph{There are guides who have no interest in
their discipline and leave their wards to their own devices.
Surprisingly, it is these guides who produce some of the most
resilient scientists, self-taught men and women, who develop great
confidence in their abilities}} [Current Science 87(2004)1319].
 \vskip 10pt
\subsection*{Doubt}
 \vskip 10pt
A researcher should provide new information to the supervisor and
avoid getting information from the supervisor.  Towards the end of 
Ph.D. course he would know much more on the topic than the supervisor.
This remarkable growth has to come through nothing but his hard work.  
There is no shortcut to success.  He should learn
and collect many information related to his work.  He should
definitely avoid embarrassing the supervisor and senior
researchers by asking doubts often.  A good supervisor or a senior
researcher does not provide answers to the questions but gives
appropriate directions to clarify the doubts.  The Nobel Laureate
Richard Feynman said: {\emph{I had no fear of doubt and
uncertainty. I don't feel frightened by not knowing things, by
being lost in a mysterious universe without any purpose.  It
doesn't frighten me.  Doubt is motivation.  It leads to discovery
and the pleasure of finding things out}}.
 \vskip 10pt
\subsection*{Complete Focus}
 \vskip 10pt
Francis Bacon said: {\emph{If a man will begin with certainties, 
he shall end in doubts.  But if he will be content to begin with 
doubts, he will end in certainties}}.
 \vskip 10pt
During the course of research, one should focus the mind mainly on
the research work. Don't allow the personal life to interfere with
research. Our life is mixed with happiness, sorrows, problems and 
difficulties.  At any stage of life how much happiness we had 
depends on how much problems we faced and how we approched each 
one of them, how we solved them and so on.  Most of the achievements 
have been made by scientists only after struggles.  If we read the 
life history of great physicists like Einstein, Galileo, Marie 
Curie, Maria Goepert Mayer and Stephen Hawkins we can notice 
that they produced great works under difficult conditions of 
their families and their own health.  Marie Curie got second Nobel Prize few years after the death of her husband Pierre Curie.  When 
Pierre Curie died their two daughters were in very childhood 
age.  Marie Curie faced severe problems in her both personal 
and academic career and inspite of all she not only won second 
Nobel Prize but also brought-up her daughter Irene Curie and her 
son-in-law Frederic Joliot to won Nobel Prize.
Maria Goepert Mayer got a permanent job with salary only in 
her 50s.  Till then she worked voluntarily without salary.  
But she came up with a work for which she was awarded Noble 
Prize.  Stephen Hawkins, when he was doing Ph.D. in his 20s 
severely affected by a neuro disease.  He is unable to walk, 
write, speak and has many health problems.  But he raised 
to the level of one of the greatest physicists of 20th century.  
He is considered as the one who knows about all physics after 
Newton.  His {\emph{Brief History of Time}} became the best 
sold book for more than three years.
 \vskip 10pt
A researcher must be clear in his thoughts.  He should know what
he has to find out.  In order to perform the work successfully the
researcher should acquire  proper training in the techniques of
research.  The training equips  the researcher with the
requirements of the task.  Further, he should be clear about his
task and possess intellectual insight.  Then only he is able to
find out the facts that would help him in his task.  Make your
research a part of your every day life.  
 \vskip 10pt
Think about your research
work in background mode, ideas will come out even when you are
seeing a movie, traveling to a place, sight-seeing and shopping.
For example, Ernst Rutherford quoted in London Times on 12
September 1933 as saying {\emph{any one who looked for a source
of power in the transformation of the atoms was talking moonshine}}.
While walking through the streets of central London after reading
this article -- as he waited for a street light at the corner of
Southampton Row -- Leo Szilard  conceived the idea of neutron
chain reaction which then led to the construction of atomic bomb.
Hans Jensen told he got the idea of the spin-orbit coupling, as 
explanation for the magic numbers which led him to won the Nobel 
Prize, one day while shaving!
 \vskip 10pt
Ted Gottfried the author of biography of Fermi said: {\emph{Scientific
research is like sports.  To score, the focus of the scientist
must be narrow and intense to the exclusion of everything else
around him.  The batter never takes his eye off the ball, the
hoopster shuts out everything but the court, the golfer always
follows through--and the scientist focuses his complete attention
on the task at hand and nothing else}}.
 \vskip 10pt
A young researcher should also have persistence, tolerance and
self-control over the unpleasant outcomes such as not getting an
expected result, not recognized by the supervisor and rejection of
a research article from a journal.  {\emph{Don't get dejected when
your paper is rejected}} -- Prof.P.R.~Subramanian.  Some times one
may complete a piece of work within a week which he might have
expected to finish it in a month time. On the other hand, at some
times one may get stuck with a particular part of the work and
unable to make a substantial progress, say, in three months. Avoid
feeling remorseful at these circumstances and maintain a high
tolerance for poor results. Remember that failure and wasted works
are also part of the research career.   Often we may be moving in long dark tunnel.  When we find a light at the end, we rejoice over it with a sense of fulfilment.
 \vskip 10pt
\subsection*{Maintaining Ties}
 \vskip 10pt
A good relationship with the supervisor is essential for several reasons.  It will greatly influence the quality and progress of your work.  Remember that maintaining an excellent relationship with another person  and working closely for a long period is not easy.  You have to maintain a fine relation with the supervisor and also with your seniors and colleagues throughout your period of association.  Backbiting and unwholesome arguments on academic matters, research work and other matters should be avoided.  The supervisor will have a thorough knowledge of the subject of research, however, at one stage you may feel that you have acquired more knowledge than him in certain topics.  But this is no reason to show-off.  You have to maintain politeness and courtesy.  
 \vskip 10pt
Professional etiquette has to be followed [\ref{warrier}].  If your are consulting another expert on some aspect of your work, it should be necessarily be with the knowledge of your supervisor. You should also take the supervisor's criticism in the right spirit and respond appropriately; there should be no reason for emotional outbursts.  
 \vskip 10pt
\subsection{\large{\bf{\emph{Getting Joy in Doing Research}}}}
To get a deep insight on the topic or the research problem a
suggestion from Dr~K.P.N.~Murthy is that {\emph{one should enjoy
doing research and approach it as an entertainment and a mode of
getting happiness}}.  In the research career one should treat
doing research as a way of life and not just a job.  In order to
achieve a goal in the research one has to work harder.  The harder
one works the happier one feels.  One need not try to conquer the
world of science.  One has to come in order to work and to find
his way. Initially one must work hard.  Getting insight in a
research topic or a research career is like a pushing a door.  It
is hard to push the door open.  But when one understand it, it is
very interesting and joyful. Enjoyment is not a goal, it is indeed
a feeling that accompanies important ongoing activity.  Gauss once
said:{\emph{It is not knowledge, but the act of learning, not
possession by the act of getting there, which grants the greatest
enjoyment}}.
 \vskip 10pt
V.V.~Raman wrote: {\emph{Associated with a selfless quest for knowledge is the excitement that comes with discovery.  An inquiring mind and the excitement of recognition have always permeated the human spirit.  The non-practitioner may find it difficult to understand the excitement of the scientific investigator who recognizes the inner workings of the world.  But this excitement has  been experienced by virtually everyone that has done even a modicum of science voluntarily and with dedication}} [V.V.~Raman, Resonance, November 2008, pp.1074-1081].
 \vskip 10pt
Eugene Wigner stated: {\emph{It has been said that the only occupations
which bring true joy and satisfaction are those of poets, artists and
scientists, and of these, the scientists are apparently the happiest}}.
Chandrasekhar pointed out that in the arts and literature quality
of work improves with age and experience while in science
generally it does not. He felt that it is because of doing science
in isolation, very narrow focus on immediate goals and
insufficient broad in interests and pursuits.  In order to
continue research even at old age one should develop the spirit of
experiencing the beauty of science.  The spirit of experiencing it
is not restricted to only the great scientists.  Chandrasekhar
said: {\emph{This is no more than the joys of creativity are restricted
to a fortunate few.  They are instead accessible to each one of us
provided we are attuned to the perspective of strangeness in the
proportion and conformity of the parts of one another and to the
whole.  And there is satisfaction also be gained from harmoniously
organizing the domain of the science with order, pattern and
coherence}}.
 \vskip 10pt
Sometimes research is reduced to a mundane activity when the sole 
intention of the researcher is to acquire a doctoral degree just 
to satisfy the administrative requirements to secure a promotion or
getting an additional increment in the salary in the academic career.
The candidate in such cases may not be motivated by a real spirit of 
inquiry, but by an eagerness to attain organizational stipulations.  
He may not enjoy the thrill at the moment of discovery of a 
new phenomenon or event or theory or explanation.  There are, 
unfortunately, ignoble enterprises that dish out ready-made Ph.Ds where
the recipients never enjoy the emotional peaks of a true researcher.  As pointed out by V.V.~Raman, the story of Archimedes running stark naked from his bathtub, screaming {\emph{Eureka!}} upon discovering a scientific principle is symbolic of the heights of joy a researcher may feel in his investigation and original scientific discovery if like a delivery of a baby: intense effort and even pain, followed by immense joy.
%
% \vskip 10pt
%\subsection{\large{\bf{\emph{Group Discussion}}}}
%
 \vskip 10pt
Professor G.~Baskaran stressed that group discussion is indeed an
important component of doing research particularly in small and
isolated institutions.  He said:  {\emph{One cannot explain the
power and usefulness of group discussions -- it has to be
experienced. When I was a student at the Indian Institute of
Science (I.I.Sc.), Bangalore, a few of us students of physics from
I.I.Sc. and National Aeronautic Laboratory were  introduced to
this joyous experience by S.K.~Rangarajan, formerly a Professor of
Chemistry, in whose house we assembled virtually every evening to
discuss such grave issues as amorphous solids and re-normalization
group. Each one of the discussants has made a mark}} (Current
Science, 75(1998)pp.1262).
 \vskip 10pt
We should able to appreciate the manner in which results are coming and notice the beauty of the methods and methodology  used get the results.

 \vskip 10pt
For a discussion on emotional factors see, for example, ref.[\ref{cs}].
 \vskip 10pt
\subsection{\large{\bf{\emph{Crucial Stage of Ph.D}}}}
The crucial period for a research scholar doing full-time Ph.D. is
the last year of the programme.  During this period one should
concentrate on completing the final work for his thesis and
writing of various chapters. Diversions to other activities should be avoided.
Further, after working about say three years and when the time has
came to consolidate the work done so far a researcher should not
start to work on an entirely new topic.  He can complete his
thesis work and then work on new topic of his interest. The woman
Nobel laureate Maria Goeppert Mayer said: {\emph{If you love
science, all you really want is to keep on working}}. 
 \vskip 10pt
Generally, a research fellowship is
for fixed period of time, it might have ended before the final
year of the Ph.D. programme.  We have noticed many scholars
converted the full-time programme into part-time and joined in a
job.  If the job is a permanent one then one can join in the job
and continue the research.  But joining in a temporary position
may highly change his research career.  This would delay the
submission of his Ph.D. thesis and he may loose the interest in
research.  There are examples with students capable of getting a
postdoctoral fellowship but failed  to even continuing the
research.  Therefore, a research scholar should have a clear plan
of what he has to do in the next few years or so.  Even if the
fellowship is not available at the finishing stage of Ph.D. thesis
we have friends and our well wishers to give financial support to
some extend.
 \vskip 10pt
\subsection{\large{\bf{\emph{The Attributes of a Research Scholar}}}}
Any researcher should be motivated by a noble goal.  Work gets the 
first, second and third priority. The attributes of a good research 
scholar may be summarized as [{\ref{warrier}}]:
\renewcommand{\labelenumi}{\theenumi}
\renewcommand{\theenumi}{$\bullet$}
\begin{enumerate}
\item
Self-confidence
\item
Dedication
 \item
Concentration
 \item
Determination
\item
Analytical mind
\item
Scientific discipline
 \item
Global outlook
\item
Innovative approach
\item
Originality
\item
Intellectual curiosity
\item
Freedom from the obsessions of clock and calendar
\item
Flexibility
\item
Keen observation
\item
Intelligence
\item
Passion for knowledge
\item
Questioning attitude
\item
Spirit of enquiry
\item
Insight
\item
Precision and accuracy
\item
Resilience to withstand temporary setbacks
\item
Persistence
\item
Patience
\item
Social skills
\item
Presentation skills
\item
Writing skills
\end{enumerate}
\section{ACTUAL INVESTIGATION}
One should aim at doing good research.  {\emph{What is good
research?}} Which universities and research institutions in your
country do the best research?  How do you distinguish the great
from a good, a black hole from an ordinary hole, a superconductor
from a normal conductor, supernova from mere stars, poles from
ordinary points, linear differential equations from nonlinear
ones?
 \vskip 10pt
To distinguish one from another we can use various quantities.
Like-wise, to identify the best from among the available, one can
use various quantities to measure the quality of them. For
example, to identify a best research the quality of the one's
research publications, number of citations of his publications,
projects completed, books published, contribution made to the
science and society, etc. can be considered.
 \vskip 10pt
Research work
\renewcommand{\labelenumi}{(\theenumi)}
\renewcommand{\theenumi}{\arabic{enumi}}
\begin{enumerate}
 \item
published in reputed international journals,
 \item
cited by other researchers working in the same or similar topic
and
 \item
which added new information to the existing knowledge on  a topic
\end{enumerate}
are generally considered as {\emph{good}}.
 \vskip 10pt
At the beginning of research career a young researcher should aim
to produce a good research, particularly, his research findings
should distinguish him from other researchers and keep him one
among the top young researchers in the nation.  In order to
encourage young researchers and motivate them to produce high
quality of research work awards are given yearly by certain
academic and research bodies in each country.  For example, in
India, Indian President Award, Indian National Science Academy
(INSA) Young Scientist Award and many other awards are given every
year.  Some Conference/Seminar organizers also provide best papers
award to young scientists.
 \vskip 10pt
%\newpage
\subsection{\large{\bf{\emph{What are the Points to be Kept
in Mind in Order to do a Good Research?}}}}
 \vskip 10pt
Actual investigation should lead to {\emph{original contribution}}
and not involve objectionable duplication.  Originality is the
basic credit point of any research.  Therefore, actual
investigation must be directed towards obtaining {\emph{novel
results}}.  A researcher should develop new ideas and obtain deep
insight into the problem in order to get novel and new results
which are the characteristics of a good research.
 \vskip 10pt
As pointed out by the physics Nobel laureate Ernst Lawrence 
{\emph{in scientific work, creative thinking demands seeing 
things not seen previously}}, or in ways not previously imagined; 
and this necessitates jumping off from {\emph{normal}} positions, 
and taking risks by departing from reality.  The difference 
between the thinking of the paranoid patient and a researcher 
comes from the latter's ability and willingness to test out 
his fantasies or grandiose conceptualizations through the 
systems of checks and balances science  has established--and 
to give up those schemes that are shown not to be valid on 
the basis of these scientific checks.  It is specifically 
because science provides such a framework of rules and 
regulations to control and set bounds to paranoid thinking 
that a researcher can feel comfortable about taking the 
paranoid leaps.  Without this structuring, the threat of 
such unrealistic, illogical and even bizarre thinking to 
overall thought and personality organization in general 
would be too great to permit the researcher the freedom 
of such fantasying.
 \vskip 10pt
Essentially, trivial analysis should not be performed.  
Recently introduced
theories, experimental techniques  and numerical algorithms have
to be used instead of outdated methods.  Before applying any
method, the researcher should familiarize with the features of the
method.  It it not worthwhile to continue in a particular
direction if the results are trivial and less informative.  If
similar problems have already been done, for instance about ten
years ago, then a researcher should not consider it as important
but could treat it as a useful exercise.   In this connection we
wish to quote the Nobel laureate Werner Heisenberg:

{\emph{If I were asked what was Christopher Columbus' greatest
achievement in discovering America, my answer would not be took
advantage of the spherical shape of the earth to get to India by
the western route -- this idea had occurred to others before him
-- or that he prepared his expedition meticulously and rigged his
ships most expertly -- that, too, others could have done equally
well.  His most remarkable feat was the decision {\bf{to leave the
known regions of the world}} and sail westward, far beyond the
point from which provisions could have gotten him back home again.
In science too it is impossible to open up new territory unless
one is prepared to leave the safe anchorage of established
doctrine and run the risk of a hazardous leap forward... However,
when it comes to entering new territory, the very structure of
scientific thought may have to be changed, and that is far more
than most men are prepared to do}}.
 \vskip 10pt
We do research by conceiving information and openings from
important research papers published by other researchers in the
topic of interest and continue in our own directions.  The work of
some other researchers might have formed the basis of our
research.  Similarly, our research outcomes should help other
researchers.  That is, the work should be such that it should
invite others to read and more importantly use it and cite it in
their research work.  Our work should lead to recognition and
respect.  It should fetch joy and benefits others and as well as
us.
 \vskip 10pt
As pointed out by Professor M.Lakshmanan, generally, {\emph{each
and every work of us may not produce novelty, but if we work
towards novelty then definitely in the course of research there
would come a fascinating and exciting breakthrough}}.
 \vskip 10pt
The researcher must remember that ideally in the course of a
research study, there should be constant interactions between
initial hypothesis, observation and theoretical concepts.  It is
exactly in this area of interaction between theoretical
orientation and observation that opportunities for originality and
creativity lie.
 \vskip 10pt
Actual work finally leads to results and conclusions of the
research undertaken.  For proper results it is necessary that
various steps of the work should be scientifically taken and
should not have any flaw.  Developed computer algorithms  must be
tested for the problems for which results are already available.
The work should be free from mistakes.  Important analysis must be
repeated in order to make sure that they are free from human
mistakes.  Professor Devanathan suggests that {\emph{a researcher
should check, recheck, cross-check, ... all the results before
submitting a research paper to a journal}}.  Before beginning to
write a part of the work done and the results obtained check and
recheck the data and the results by repeating the experiment,
rerunning the programs and going through the theoretical
derivations and arguments.
 \vskip 10pt
When analysing the data, appropriate statistical tools have to be
employed.  The number of data used, units of the data, error bars
and other necessary details must be noted in the graphs.  As many
statistical tools as possible should be used.  Appropriate curve
fitting can be done.  Necessary interpretations on the results of
statistical analysis have to be made.
 \vskip 10pt
In the case of development or modification of a theory and
proposal of a new method the assumptions made, basic idea, and
calculations should be clearly stated and  analyzed.  Various
special cases of the theory or method must be identified.  The
validity, efficiency and applicability of it must be demonstrated
with examples.  Merits and demerits have to be identified.
Comparison of the proposed method with the already existing and
widely used similar methods should be performed.
 \vskip 10pt
In any experimental work, mere measurement of certain quantities
is not enough.  The interpretation of the kind of data observed
and  explanation for the particular pattern must be made. On the
basis of interpretation general principles underlying the process
can be formulated.  One has to check whether the generalizations
are universal and true under different conditions.
 \vskip 10pt
Some common errors made in research are [\ref{camden}]
\renewcommand{\labelenumi}{(\theenumi)}
\renewcommand{\theenumi}{\arabic{enumi}}
\begin{enumerate}
 \item
Selective observation
 \item
Inaccurate observation
 \item
Over-generalization
 \item
Made-up information
 \item
Ex post facto hypothesizing
 \item
Illogical reasoning
 \item
Ego involvement in understanding
 \item
Premature closure of inquiry
 \item
 Mystification
\end{enumerate}
For a very interesting discussion on the above aspects with
examples refer to the ref.[\ref{camden}]
%
%\newpage
\section{RESULTS AND CONCLUSION}
The next step after performing the actual research work on the
chosen problem is preparation of results and conclusion of the
performed work.  Predictions, results and conclusion are ultimate
goals of the research performed.
 \vskip 10pt
There are two indispensable rules of modern research.
The freedom of creative imagination necessarily subjected to
rigorous experimentation.  In the beginning any experimental
research on a specific subject, imagination should give wings
to the thought.  At the time of concluding and interpreting the
facts that were collected observation, the imagination should be
dominated and prevailed over by concrete results of experiments. 
We should analyse cause and effect. We should pay attention to minute details also.  In fact keenness in observation is the hallmark of any scientific research.
 \vskip 10pt
Proper interpretations of the results must be made.
{\emph{Interpretation refers to the task of drawing inferences
from the actual research work}}.  It also means drawing of
conclusion. Conclusion is based on the study performed. It would
bring out relations and processes that underlie the findings. The
utility of the outcome of the research greatly lie on proper
interpretations and is the hardest part of solving a scientific
problem. Interpretation of results is important because it
\renewcommand{\labelenumi}{(\theenumi)}
\renewcommand{\theenumi}{\arabic{enumi}}
\begin{enumerate}
 \item
links the present work to the previous,
 \item
leads to identification of future problems,
 \item
opens new avenues of intellectual adventure and stimulates the
quest for more knowledge,
 \item
makes others understand the significance of the research findings
and
 \item
often suggests a possible experimental verification.
\end{enumerate}
 \vskip 10pt
The basic rule in preparing results and conclusion is to give all
the evidences relevant to the research problem and its solution. A
bare statement of the findings are not enough. Their implications
must be pointed out.  Discuss your answers to the following
questions with experts:
\renewcommand{\labelenumi}{(\theenumi)}
\renewcommand{\theenumi}{\arabic{enumi}}
\begin{enumerate}
 \item
Are the supporting evidences sufficient?, and if not, What to do?
\item
How many pieces of evidence are required?  Instead of producing
all, is it possible to restrict to one or two pieces of evidence?
If so, what are they? and
\item
Why are they sufficient?
\end{enumerate}
and so on. Such directions can help us minimize work and the
quantity of presentation of the report.  Do not rely on a bogus
evidence which would increase the chances of errors. The
investigator has to give suggestions. These should be practical
and based on logic, reasoning and fact. The suggestions should be
such that they can be actually implemented.
 \vskip 10pt
According to Feynman (Surely you're Joking, Mr.Feynman!) if we are
doing an experiment, we should report everything that we think
might make it invalid--not only what we think is right about it;
other causes that could possibly explain our results; and things
we thought of that we have eliminated by some other experiment,
and how they worked--to make sure that other fellow can tell they
have been eliminated.  Further, details that could throw doubt on
the given interpretation must be included, if such is known.  You
must do the best you can--if you know anything at all wrong--to
explain it.  If you make a theory, for example, and advertise it,
or put it out, then you must also put down all the facts that
disagree with it, as well as those that agree with it.
 \vskip 5pt
The researcher should not be in hurry while preparing the results
and conclusion.  After preparing them the researcher may ask the
following questions:
\renewcommand{\labelenumi}{(\theenumi)}
\renewcommand{\theenumi}{\arabic{enumi}}
\begin{enumerate}
 \item
Are the quantitative and qualitative analysis performed
{\emph{adequate}} for the conclusion drawn?
 \item
Are the results and conclusion {\emph{general}}?
 \item
Are the results and conclusion {\emph{valid only for the
particular situation}} considered in the present work?
 \item
Is the conclusion {\emph{too broad}} considering the analysis
performed?
 \item
Is any evidence which {\emph{weaken}} the conclusion omitted?
\end{enumerate}
\noindent The results and conclusion prepared can be revised based
on the answers to the above questions.
 \vskip 10pt
Each and every statement made in the results and conclusion
sections must be based on evidence obtained from theoretical or
experimental analysis.  Baseless statements should never be made.
 \vskip 10pt
While doing research particularly experiments, one may land up with an unexpected result or a finding contrary to the underlying theory.  Such an observation should not be ignored blindly.  It may be pursued to some extent to check whether it yields some useful result.  As noted by Warrier [\ref{warrier}] the history of science records such fortuitous breaks that led to many classic discoveries. 
 \vskip 10pt
Never yield to the temptation of fabrication of results and interpretation.  Plagiarism in the form of copying data or findings from others' report without acknowledging the source will make you in trouble.  Don't attempt to report the already reported findings of others  as yours.  Citing the original sources  actually enhances the credibility of your work.  
 \vskip 10pt
One should note that the observations, claims and the conclusions drawn in a research report are subjected to a criticism by the experts in the concerned field.  Therefore, the researchers have to think twice before presenting the outcomes of the research in a journal and in a scientific meeting.
 \vskip 10pt
 \hrule
 \vskip 5pt
\noindent{\bf{Assignment:}}
\renewcommand{\labelenumi}{(\theenumi)}
\renewcommand{\theenumi}{\arabic{enumi}}
\begin{enumerate}
 \addtocounter{enumi}{8}
 \item
For each of the following topics write at least two questions, the
answers to which must be available in the respective topics.  For
example, for the topic, {\emph{introduction}}, a relevant
question is {\emph{Why am I doing it?}}.

(i) Introduction, (ii) Review of a research topic, (iii)
Methodology, (iv) Research design, (v) Results, (vi) Discussion
and (vii) Conclusion.
\end{enumerate}
 \hrule
 \vskip 5pt
\sectionmark{PRESENTING A SCIENTIFIC SEMINAR--ORAL REPORT}
\section{PRESENTING A SCIENTIFIC SEMINAR-ORAL REPORT}
\sectionmark{PRESENTING A SCIENTIFIC SEMINAR--ORAL REPORT}
\subsection{\large{\bf{\emph{What is an Oral Report?   What are the
Importance of an Oral Report?}}}}
 \vskip 2pt
Presentation of one's research work in a scientific meeting is an
{\emph{oral report}}.  Scientific meetings include conference,
seminar, symposium, workshop, departmental weekly seminar, etc.
 \vskip 10pt
Researchers in certain research institutions not only discuss
their own work but also have discussions on very recently reported
work of other scientists.
 \vskip 10pt
An oral report provides a bridge between the researcher and
audience and offers greater scope to the researcher for explaining
the actual work performed, its outcome and significance. It also
leads to a better understanding of the findings and their
implications. In an oral report, the researcher can present the
results and interpretations which are not clearly understood by
him and may request the experts in the audience to give their
opinions and suggestions.  Oral reporting at a conference or a
seminar requires more elaborate preparation than the written
report.
 \vskip 10pt
A Nobel Prize winner Paul Dirac said: {\emph{A person first gets
a new idea and he wonders very much whether this idea will be
right or wrong.  He is very anxious about it, and any feature in
the new idea which differs from the old established ideas is a
source of anxiety to him.  Whereas  some one else who hears about
this work and talks it up doesn't have this anxiety, an anxiety to
preserve the correctness of the basic idea at all costs, and
without having this anxiety he is not so disturbed by the
contradiction and is able to face up to it and see what it really
means.}}
\subsection{\large{\bf{\emph{Points to be Remembered in Preparing
an Oral Report}}}}
Before starting the preparation of an oral report, an outline can
be drawn based on the time duration of the report and the quality
of the audience.  Departmental seminar is usually 45 minutes
duration.  In other meetings time duration is fixed by the
organizer based on the number of days of the meeting, number of
speakers and the status of a speaker.
 \vskip 10pt
For a long time report, that is, 45--60 minute presentation, one
may have enough time to
\renewcommand{\labelenumi}{(\theenumi)}
\renewcommand{\theenumi}{\arabic{enumi}}
\begin{enumerate}
 \item
introduce the topic,
 \item
discuss the definition of the problem,
 \item
describe the method and technique employed,
 \item
give technical details, and
 \item
present results and conclusion.
\end{enumerate}
Consequently, these aspects can be prepared in detail.
 \vskip 10pt
For a 15--30 minute, oral presentation one cannot find enough time
to discuss complete details of the work.  In this case less
informative materials must be dropped.  Methods and techniques used
can be presented very briefly without going into technical
details.  Much time should be reserved for results, conclusion and
further directions.
 \vskip 10pt
Prepare a write-up of the oral presentation.  It is a good and very 
helpful practice to write the talk before presenting it orally.  
Then evaluate the written material. Ask:
\renewcommand{\labelenumi}{(\theenumi)}
\renewcommand{\theenumi}{\arabic{enumi}}
\begin{enumerate}
\item
{\emph{Why should the audience listen to your presentation?}}
\item
{\emph{Is the presentation match with the standard of the
audience?}}
\end{enumerate}
Revise the presentation until you get convincing answer to the
above two questions.  Make sure that your objective would convince the audience that you have done your job well, your methodology is sound and the findings are useful.
 \vskip 10pt
The success of a presentation lies in making it {\emph{long enough to 
cover the topic and short enough to arouse curiosity}}.
Oral presentation can be made effective and attractive by using
modern visual devises, power-points, slides and transparency
sheets.  Title of the report, author's name, plan of the
presentation, very important content of it and conclusion can be
printed in the slides or sheets possibly point by point with bold
and sufficiently large size letters.  Merely reading out measured 
or computed data will never catch the attention of the audience.  
They may be displayed in the form of histograms.
Important formulas,
equations, tables, figures and photographs can be prepared using
transparency sheets or slides.  Slides and transparency sheets
should not contain running matters.  {\emph{Researcher should not
simply read the content in the sheets}}. That is, the descriptive
portion of the report should not be prepared on the sheets.  An
abstract or a short write-up of the presentation may be circulated
to the participants of the meeting.  Sophisticated softwares
developed for preparing the text on transparency sheets/slides are
available in internet and can be freely downloaded. In order to
make the presentation, more lively, the researcher could use
multimedia.  Nowadays, the use of {\emph{power-point}} of
Microsoft Windows is common. It is an easy and compact utility
software especially for preparing classroom presentations. The
following are the web sites from which one could download the
software at free of cost:
 \vskip 3pt
 http://www.office.microsoft.com/downloads
 \vskip 1pt
 http://www.lb.com/download-free-power-point-presen\-tation.org
 \vskip 10pt
One could use the audio aspects also to facilitate his
presentation in a better way.  While presenting the topic, the
researcher should strictly follow the class room teaching
methodology.  For example, one should allow interaction; don't
forget to modulate the voice as and when required and don't
violate the time frame.  Logical continuity is another key aspect.  
Move from the simple to the complex, from the known to the unknown.  
Your statements should sound sensible and reasonable.  Do not speak
too fast and compromise on clarity, or speak too slowly and bore the 
audience.  Make the session interactive by posing questions.
As pointed out by Warrier, there is a dictum to be followed in good speeches:
{\emph{First tell them what you are going to tell.  Next, you tell them. 
Then tell them what you told them}}.
 \vskip 10pt
The most crucial part is the actual presentation in front of the
listeners.  Stage fright is the bane of most presenters.  There is 
no shortcut to overcome this fear.  One has to practice, practice 
and practice.  You can improve your presentation skill by getting the
feedback after you have spoken. Avoid repeated use of words or phrases 
such as ``well", ``your see", ``you know", ``I mean", ``I think", ``that is" and ``basically".
You should also concentrate on your body language.  Smooth movements 
of limbs, an occasional smile and pleasing manners would endear us to 
the listeners.  Look relaxed and comfortable.  Eye contact should be 
maintained with the audience.  Let your eyes move uniformly around 
the entire audience and not focus on any particular person or a particular part of the audience.
 \vskip 10pt
One or two rehearsals of the report in the presence of colleagues,
supervisor and collaborators can be exercised in order to
\renewcommand{\labelenumi}{(\theenumi)}
\renewcommand{\theenumi}{\arabic{enumi}}
\begin{enumerate}
 \item
complete the presentation within the allotted time,
 \item
improve the quality of presentation and
 \item
maintain the fluency of the presentation.
\end{enumerate}
During a long presentation, the speaker can stop the presentation
at various stages, seek comments and questions from the audience
and then proceed. This will make the presentation attractive,
interesting and also allow the audience to clarify their doubts so
that they can follow the work.  Your clarifications should be clear and convincing.  Never go into arguments.  Your confidence gets boosted, if you approach the situation with a feeling that the expert panel is there to help you.  An important
point is to consider the tone to adopt so that you sound genuine.
 \vskip 10pt
For more details on how to make your presentation more effective, see ref.[\ref{warpre}].
\section{ART OF WRITING A RESEARCH PAPER AND THESIS}
\subsection{\large{\bf{\emph{What is a Research Report?}}}}
{\emph{Research reporting}} is an oral or a written presentation
of important and useful aspects of the research work done.
Scientific writing, a thesis or a paper, is intended to present
clearly the purpose and outcome of a specific research
investigation. It is the last but a major part of the research
study. A report helps the researcher get feedback from other
researchers and experts working in the same field.  It also
evaluates the success and originality of the researcher's work.
{\emph{Without a report, a research study is incomplete and of no
use}}.  A report essentially conveys the outcome of a research
work to interested persons. Brilliant work and most striking
findings are of little value if they are not effectively
communicated to the scientific world.  As pointed out by Eli Maor,
{\emph{in academic matters the iron rule is publish or perish}}.
Some times  delaying a publication of a result one would lose his
claim.
%
%\newpage
\subsection{\large{\bf{\emph{What are Research Paper or Article 
 and Ph.D Thesis or Dissertation?}}}}
A research paper is a report published in a journal or magazine or
conference proceedings, etc.  Whereas a Ph.D. dissertation is a
report of the entire work done by a researcher to a university or
an institution for the award of the degree of doctor of
philosophy. The central element of a Ph.D. education is the 
doctoral thesis, that is, the Ph.D. dissertation.  It is an 
apprenticeship in {\emph{how to do research}} and forms the 
unique part of Ph.D. degree course.
A Ph.D. dissertation is a lengthy, original and
substantial document.  It should contain original contributions.
Essentially, the {\emph{role of a Ph.D. dissertation is to
demonstrate the research person's original thinking and
contribution to the topic of research}}.  It should also clearly
point out the research competence of the researcher in his
research field.  M.Phil. dissertation is designed as a practice
for Ph.D. thesis.  It will help the researcher learn and
understand the present status of the topic and make him capable of
working at the Ph.D. level. The work done for an M.Phil.
dissertation need not be publishable in journals.
\subsection{\large{\bf{\emph{Why Should a Researcher Report  
his Findings?}}}}
{\emph{Every research investigation is carried out with certain
objectives}}.  The outcome of a research work may add new
information to a theory or may have technological applications.
Sometimes the researcher may not be aware of the theoretical
development on practical applications.  His research results may
be useful to another research problem.  Some other researchers may
be working or planning to work on the same or similar type of
research work.  Several researchers doing same research work is a
waste of time unless the solution of the problem is needed very
urgently and is of great use.  Repetition of a work should be
avoided by the research community as much as possible.  Unless a
researcher reports his work to the world, the {\emph{correctness,
validity and originality}} of the work is under a question mark. The
outcome of a research work will become known to the scientific
community only through publications. G.~Madhavan writes
{\emph{a piece of research not published is as good as not carried out}}.
 \vskip 10pt
In view of the above, it is important to report a work in an
appropriate journal or magazine and in scientific meetings like
conferences, seminars and symposia. Identify possible publications
of your research findings after making a considerable progress on
a research problem.  Don't be confined with a mere Ph.D. degree.
\subsection{\large{\bf{\emph{Characteristics of a Good Report}}}}
A good report results from slow, pain taking and accurate
inductive work.  To attract a reader, the reading matter of a
report should be clear and interesting.  It should not be obscure
and dull.  The write-up should be logical, clear and concise. The
physicist Cyril Isenberg (University of Kent) writes: {\emph{One
has to present the work like a meal in a visually attractive and
palatable way. It must be easily digested and, one hopes, long
remembered.  A paper that is difficult to understand and is not
presented in a logical way, relying heavily on results in other
papers, is like a dry biscuit, with little sustaining value and
even less taste.  It will not be digested and will be left unread
or abandoned}}.  The basic quality or characteristics of a good
scientific report/paper and thesis are the following:
\renewcommand{\labelenumi}{(\theenumi)}
\renewcommand{\theenumi}{\arabic{enumi}}
\begin{enumerate}
 \item
good presentation
 \item
good organization of various chapters/sections
 \item
accuracy
 \item
clarity
 \item
free from contradictions and confusion.
\end{enumerate}
Further, a Ph.D. dissertation should be a formal and should have
high level of scholarship.
\section{OUTLINE OF A REPORT}
\noindent{\large{\bf{\emph{What are the considerations to be kept
in mind while preparing a report?}}}}
\renewcommand{\labelenumi}{(\theenumi)}
\renewcommand{\theenumi}{\arabic{enumi}}
\begin{enumerate}
 \item
First, an outline of a report has to be prepared.
 \item
A sketch of what information to be conveyed must be made.
 \item
Then, one can write down various topics, subtopics to be
considered and what material to be presented in them.
 \item
The sentences which are to be expanded, reworded and verified for
its validity can be marked.
\end{enumerate}
The outline of the report helps us concentrate on
\renewcommand{\labelenumi}{(\theenumi)}
\renewcommand{\theenumi}{\roman{enumi}}
\begin{enumerate}
 \item
what is to be presented,
 \item
logical relationships between different parts of the report,
 \item
smooth flow of the content and
 \item
continuity in the presentation.
\end{enumerate}
The outline can be discussed with the guide, collaborators,
colleagues  and experts in local area.  Based on their comments
the structure of the report can be modified.
 \vskip 10pt
A three stage preparation of a report is generally done by
researchers.  They are:
\renewcommand{\labelenumi}{(\theenumi)}
\renewcommand{\theenumi}{\arabic{enumi}}
\begin{enumerate}
 \item
First draft -- {\emph{Rough draft}}.
 \item
Second draft -- {\emph{Rewriting and polishing of the rough
draft}}.
 \item
Third draft -- {\emph{Writing the final draft}}.
\end{enumerate}
\subsection{\large{\bf{\emph{First Draft}}}}
In this stage a researcher can write
\renewcommand{\labelenumi}{(\theenumi)}
\renewcommand{\theenumi}{\arabic{enumi}}
\begin{enumerate}
 \item
what has been done in the research study,
 \item
procedure, method, theory and technique applied,
 \item
technical difficulties faced and how they were overcame,
 \item
broad findings and
 \item
concluding remarks.
\end{enumerate}
Tables and charts can be typeset using computer and kept
separately in order to avoid rewriting them.  Conclusion should be
precise, clear and objective.  Further directions may be pointed
out.
 \vskip 10pt
Since a research paper is identified by its title it should be
brief and not more than above 10-15 words.  A subject index of a
paper is primarily based on the words in the title.  Therefore,
few key words which are helpful to classify the paper can be
included appropriately in the title.
 \vskip 10pt
How does a reader decide whether to read the content of a paper or
not?    Abstract serves the purpose.  By reading the abstract a
reader would decide whether the content of the paper is useful to
him.  Therefore, the abstract should have positive information
about the content of the paper and summary of the work reported in
it.  Further, if the abstract has final results and main
conclusion of the paper then a reader who has a general interest
in the subject can know the outcome of the paper without reading
the entire text by referring the abstract itself.
\subsection{\large{\bf{\emph{Second Draft}}}}
This is the most important and difficult part of the writing.
Extreme care must be taken in writing this draft.  
One convenient approach is to edit the draft imagining 
that someone else has written it.
 \vskip 10pt
Unclear points, jargons, weakness of the report have to be identified and revised.
Over-generalization of outcomes should be avoided.  For example,
Hermitian operators have real eigenvalues.  Generalizing it as
eigenvalues of operators are real or concluding that to have real
eigenvalues, operators should be Hermitian are incorrect.
Similarly, complex analytic functions satisfy Cauchy--Riemann
conditions.  It does not mean that functions satisfying
Cauchy--Riemann conditions should be analytic. How do you avoid
over-generalization?  For some details see, for example, ref.[\ref{cs}].
If you have introduced any graphics and tables, discuss it in the text.
 \vskip 10pt
Attention must be paid to the arguments made, logical flow of work
presented, the quality of supporting evidences and conclusion
drawn.  Do these in each  chapter. Don't do the entire second
stage at a single stretch.  Give sufficient time between revisions
of two consecutive chapters. During the break time think over the
revision made in the previous chapter or section.
 \vskip 10pt
More importantly, grammar must be checked. A careful spell check
must be made.  Use simple words as far as possible.  Indecisive
words such as perhaps, somewhat, rather, etc. should be avoided.
Usage of some particular words repeatedly, for example, `very',
`extraordinary', `invariably' should be avoided.  Expressions such
as `it seems', `there may be', `since', `putting', etc. should be
replaced by appropriate equivalent words.
 \vskip 10pt
Style, presentation and grammar can be improved by asking your
friends, colleagues to read and give their critical comments,
suggestions and correct English grammar.
 \vskip 10pt
In some universities the report is first read by an English
teacher.  He will correct the grammar and give suggestions.  After
this only a researcher can submit the thesis.
 \vskip 10pt
Complicated and lengthy sentences have to be rewritten and broken.
Similar sentences or sentences conveying same information must be
eliminated.  Check whether the words used clearly convey exactly
the meaning intended.
 \vskip 10pt
S.~Chandrasekhar said: {\emph{I always sought to present my findings in
as elegant, even literary, a form as possible.  I select some
writers in order to learn.  For example, I read Henry James or
Virginia Woolf, and I don't simply read the text as a novel;  I
see how they construct sentences, how they construct paragraphs,
how one paragraph goes into another and so on.}} (J.~Horgan, Current
Science, 67~(1994)~pp.500-01).
 \vskip 10pt
B.S.~Warrier says: {\emph{There is one clear distinction between 
an article and a thesis in use of vocabulary.  You may find 
that authors of popular writing make use of synonyms 
to kill monotony.  For example, instead of repeating the 
word teaching, they may use instruction, coaching, training, 
schooling, tutoring and education, ignoring the fine 
differences in the shades of meaning.}}. 
(The Hindu, 20 November 2006 p.6 of Education Plus).
\vskip 10pt
The conclusion may carry something more than a 
repetition of the findings indicated elsewhere.  
All conclusions should be directly related to the research.  Perhaps 
the conclusion could reveal a special insight of yours, throwing up a possibility of the findings being applied to a different  
situation or even different discipline.
Proper references of related work should be included. Trivial
matters and obvious conclusion should not be included and if there
are such sentences then  they should be dropped.
\subsection{\large{\bf{\emph{Third Draft}}}}
This is the last stage.  In this stage, one can concentrate on
{\emph{final touches and finishing}}.  This should be in the
direction of making the report weighty, authoritative, attractive
and convincing.  Similar words and format should be avoided in
successive sentences.  Make sure that the script clearly shows the
originality of the author and importance of the outcome of the
study performed.
  \vskip 10pt
In all the three stages of report preparation one should follow a
proper style of writing.  Use clear and unadorned English
appropriate for the readers. One has to be aware of to whom the
research report is intended.  {\emph{The report is not for the
supervisor}}. It is better to avoid the use of personal pronoun. Use
of ``I" and ``the author" should be avoided.  Some supervisors
like to use ``we".  For an interesting fun about the usage of ``I"
and ``we"  see p.106 of {\emph{Why are things the way they are?}} by
G.~Venkataraman (University Press, Hyderabad, 1992).
 \vskip 10pt
Both active and passive voice should be used wherever necessary or
appropriate.  However, when using them one should check whether
the meaning is strictly correct.  For example, when writing {\emph{The
experimental results agree with the theory}}  we must check whether
we are strengthening the experimental result or the theory.  Care
must be taken in using present and past tenses. Use past tense to
describe the data collection and work done by others and you.  For
interpretation, assessments and discussions present tense is
appropriate.
\vskip 10pt
Since a research article  is to be read by experts in the field, avoid
expressions such as ``In other words" and ``To put it in a simpler way."
Experts may not take kindly to such phrases, as they feel that 
they are competent to grasp even complex ideas stated in difficult idiom.
 \vskip 10pt
Between various stages it is advisable to give gap of few days so
that you can leisurely think of the manuscript and record how to
revise it.  This will avoid unnecessary tension and half-hearted
write up.
 \vskip 10pt
{\emph{How do you ensure that your paper will be well received by readers?}}
 Some of the suggestions are:
\renewcommand{\labelenumi}{\theenumi}
\renewcommand{\theenumi}{$\bullet${}}
\begin{enumerate}
 \item
Before sending it to a journal, authors can ask their colleagues
and friends working in the same field to read it and comment on
it.
\item
The paper can be given to colleagues who are not familiar with the
topic so that it can be attracted by non-specialists in the field.
\item
Don't feel offended if the colleagues suggest  drastic changes.
\item
Think back to couple of papers which have made an impression on
you and are still long remembered.  Note down and analyse the style,
presentation and other features that have impressed you.
\end{enumerate}

As pointed  out by David Hilbert
the presentation and understanding of the outcome of an investigation is not to be considered complete until you have made it so clear that you can explain it to the first man whom you meet on the street.  This clarity and ease of comprehension is so important.
\sectionmark{LAYOUT OF A PH.D. THESIS / M.PHIL. DISSERTATION}
\section{LAYOUT OF A RESEARCH REPORT /
 PH.D. THESIS / M.PHIL. DISSERTATION}
\sectionmark{LAYOUT OF A PH.D. THESIS / M.PHIL. DISSERTATION}
The layout of a research report is the list of various parts of
the report/thesis.  Generally, a research report should consist of
the following three components:
\renewcommand{\labelenumi}{(\theenumi)}
\renewcommand{\theenumi}{\arabic{enumi}}
\begin{enumerate}
 \item
Preliminary pages
 \item
Main text
 \item
End matters
\end{enumerate}
\subsection{Preliminary Pages}
{\emph{Preliminary pages include title of the report,
acknowledgment, certificate page, list of publications and table
of contents}}.  Acknowledgments are written to thank those who
have helped the researcher during their course of investigation.
For a book it is in the form of preface or forward.
Acknowledgment should be brief, simple, modest and given only to
substantial assistance provided by the guide, head of the
department, staff of the department, agencies which provided
financial support, collaborators and institutions where part of
the work has been carried out.  Acknowledgments made for routine
participation by members of the researcher's family, librarian,
friends, clerical helpers and god are normally considered
superfluous. Acknowledgment should be made at the time of public
viva-voce also.  There is a chance for a researcher to forget to
say acknowledgment at the end of an oral presentation.  To avoid this
he may do it at the beginning of the presentation.  
 \vskip 10pt
Declaration in the certificate page by the scholar is generally done 
using phrases such as ``I hereby declare that this submission is my 
own work and that, to the best of my knowledge and belief, 
it contains no material previously published or 
written by another person nor material which to a substantial 
extent has been accepted for the award of any other degree 
or diploma of any university or institute of higher learning, 
except where due acknowledgment has been made in the text.
\vskip 10pt
{\emph{Every research report should have an abstract}}.  It is a
necessary part of any scientific and nonscientific research
report. In a research article it appears next to the author's name
and affiliation.  In the case of Ph.D. thesis, before its
submission an elaborated abstract of the thesis called
{\emph{synopsis}} has to be submitted to the institution where
registration for Ph.D. degree is made.  Abstract and synopsis
convey the essence and brief details about the report. It should
contain a very short statement of the problem, methodology and
procedures adapted in the work and results of the study in a very
condensed form.  {\emph{The abstract can act as a tool to control
the flow of ideas in the thesis}}.  It can help you link in a
logical way the reasons for the research and aims of the work.  It
should contain answers to the questions: What was done in the
project? Why is it of interest?  How was it done? What were the
outcomes of the work done?  What is the significance of the
results? One should emphasize the original contribution in the
abstract. The abstract of a Ph.D. thesis will be about three or
four pages.
  \vskip 10pt
{\emph{Table of contents gives title of the chapters, section
headings, title of appendices and their page numbers}}.  All the preliminary pages should be numbered with lower-case roman
numbers.
\subsection{Main Text}
The main text presents the details of the research work and
results. This part of the thesis should provide the following,
about the research work:
\renewcommand{\labelenumi}{(\theenumi)}
\renewcommand{\theenumi}{\arabic{enumi}}
\begin{enumerate}
 \item
Introduction.
 \item
Actual research work performed and the findings.
 \item
Summary and conclusion.
\end{enumerate}
\subsubsection{\large{\bf{\emph{Introduction}}}}
The purpose of the introduction is to give a brief outline of the
field of research.  In this part one can bring clearly the
importance of the field and the current status of it.  It should
contain an overview of the problem, its importance, statements
about the hypothesis or specific questions to be explored.  This
is followed by a preview of the scheme of the following chapters,
that is an outline of plan of the work. Here, aim of each of the
chapters and their contents can be briefly stated.  Related and
relevant work done by others must be pointed out.  Various
concepts and definitions of scientific and technical terms
necessary for understanding the research work undertaken are to be
defined and explained.  Details of statistical tools or quantities
used in the study can be given in a separate chapter.
 \vskip 10pt
Irrelevant and less informative materials need not be presented.
For example, regular and irregular behaviour of solution of a
system or differential equation can be characterized by
calculating the statistical tools such as Lyapunov exponents,
correlation function, correlation dimension, power spectrum,
periodicity of the solution and probability distribution.  If the
power spectrum is not used in a research work then there is no
need to discuss in detail the systematic way of calculating it.
Similarly, suppose the effect of noise in a theoretical model
equation is studied by including, say, Gaussian random numbers in
the simulation.  There are many methods available to generate
Gaussian random numbers.  If the Box--Muller method is used then
it can be described.  In this case describing other methods, for
example, rejection technique is redundant to the present thesis
report.  The theory and experimental set up used should be clearly
described with proper references. Define the technical terms used
in the dissertation either by a reference to a previously
published definition or by a precise definition.  Such a
definition should be given only once in the report.
 \vskip 10pt
The introductory chapter(s) should be prepared in such a way that
it should interest the reader in the subject matter of research.
It should not be aimless, confused and lacking in precision.
Introductory part may contain one or two chapters.
  \vskip 10pt
To be precise, the introductory part should cover the following
aspects:
\renewcommand{\labelenumi}{(\theenumi)}
\renewcommand{\theenumi}{\arabic{enumi}}
\begin{enumerate}
 \item
Features of the topic
 \item
Present status of the field
 \item
Some unsolved problems
 \item
Statement of the problem undertaken
 \item
Importance and justification of the present problem
 \item
Preview of the scheme  of the following chapters and their
interrelationship definition of various scientific terms used
 \item
Methodology used
\end{enumerate}
\subsubsection{\large{\bf{\emph{Actual Research Work}}}}
This is the heart of the research report/thesis.  The actual
research work undertaken, difficulties faced, technical details,
results, conclusion and future direction form the main part of
this portion.  This part can be presented in a few chapters.  Each
chapter should contain introduction, research work, results and
conclusion.  Materials should be organized systematically and
presented under appropriate headings and subheadings.  First,
write the chapters that describe your actual research work.  After
this, prepare the conclusion and introduction parts.  When writing
the actual work collect the terms and note down the matter which
are to be defined and described in the introduction.
 \vskip 10pt
As Professor P.R.~Subramanian points out, {\emph{for preparing the
Ph.D. thesis report one should not simply copy word by word from
his research articles}}.  Even if the content of the thesis is the
work reported in his research publications, the student should
reword the material without changing the meaning, give much more
details, explanations, suggestions and possibly a better
reorganization of the content.
  \vskip 10pt
Wherever possible, the results should be presented in the form of
figures, illustrations and tables. They can make the report quite
attractive.  Tables should be as precise as possible.  All the
figures should clearly specify the variables of the axes, units
used and other necessary information.  Figure caption should not
be a reproduction of sentences of the text. It must clearly state
what it is. Figures should be clearly explained in the text.  Data
should be fitted to an appropriate mathematical expression.
Nowadays, sophisticated softwares are available for curve fitting.
After making a curve fit or plotting a set of data, proper
explanation for observed variation of the data should be given.  A
set of data measurement without any analysis and discussion is of
no use.
 \vskip 10pt
Arguments may be conveniently presented as a series of numbered 
or bulleted points, rather than as one chunk in a crowded paragraph.  
Mention  further unexplored areas, which future researchers may conquer.
  \vskip 10pt
Extreme  care must be  taken in type setting  mathematical
equations, variables and parameters involved in the study. Italic
or Greek letters or mathematical symbols can be used for variables
and parameters. For example, x or X should not be used as a
variable name.  The correct usage is $x$ or $X$ (or typeset in
italics). All the equations should be centered and numbered.
Vectors should be clearly specified by an arrow over the name or
by bold face name.  Equations should not be repeated.
 \vskip 10pt
Jokes or puns should not find a place in the report. Use
``correct" or ``incorrect" to refer to the results of others.
Don't use the words ``bad", ``terrible" and ``stupid".  Avoid use
of ``today", ``modern times", ``soon", ``seems", ``in terms of",
``based on", ``lots of", ``type of", ``something like", ``just
about", ``number of", ``probably", ``obviously", ``along with",
``you", ``I", ``hopefully" and ``may".  There is no need to
mention the circumstances in which the results are obtained.
\vskip 10pt
An error often made is wasting valuable time for the physical 
embellishment of the document beyond a point, without 
paying careful attention to the correctness and accuracy of the content.  
Even a couple of typos can give the  impression that you have failed to pay 
adequate attention to detail.  Errors in the spelling or technical or general 
words show in the poor light an otherwise worthy thesis that tells a vital story.
 \vskip 10pt
 \hrule
 \vskip 5pt
\noindent{\bf{Assignment:}}
\renewcommand{\labelenumi}{(\theenumi)}
\renewcommand{\theenumi}{\arabic{enumi}}
\begin{enumerate}
 \addtocounter{enumi}{9}
 \item
Reword/rephrase the following and give the reason for the change:
\renewcommand{\theenumi}{\roman{enumi}}
\begin{enumerate}
 \item
Dinesh and Geethan [1] reported that ...
\item
The following algorithm represents a major breakthrough ....
\item
Even though the above method is not earthshaking ....
\item
Geethan and I obtained ....
\item
There is a method to calculate ....
\item
The program will use the data after it stored them  to a CD ...
\item
The method is started by calculating the value of $\delta$ ....
\end{enumerate}
\end{enumerate}
 \vskip 5pt
 \hrule
\subsubsection{\large{\bf{\emph{Conclusion}}}}
At the end of each chapter (except in the introductory chapter(s)),
one can place a brief summary of
the outcome of the work presented in that chapter under the
heading conclusion. They should be clear and precise.
 \vskip 10pt
The relevant questions which are still not answered and new
questions raised by the work of the present chapter have to be
mentioned. Whether the answers to the questions are obtained or
not, if obtained in which chapter(s) they are presented should be
specified.  Mention possible future research. It is important to
make a connection between two consecutive chapters either at the
end of the first or at the beginning of the second.
 \vskip 10pt
Chapters should not look like reports of isolated work.  There
should be a link between consecutive chapters and the link should
be clearly brought out.
\subsection{End Matters}
The end part of the report generally consists of references,
appendices, computer programs (if they are not easy to develop)
and copies of research publications that came out from the
research work done.
\subsubsection{\large{\bf{\emph{Appendices}}}}
Appendices are supplementary contents which are not placed in the
main report in order to keep the continuity of the discussion;
however, they are relevant for understanding the particular part
of the report.  An appendix may present
\renewcommand{\labelenumi}{(\theenumi)}
\renewcommand{\theenumi}{\arabic{enumi}}
\begin{enumerate}
 \item
a brief summary of a theory or a numerical method used which can
be found elsewhere,
 \item
a lengthy mathematical derivation or a large set of equations,
 \item
technical details and
 \item
a list of values of constants and parameters used in the work.
\end{enumerate}
Appendices can be placed at the end of report after references.
They should be numbered by capital alphabets.
\subsubsection{\large{\bf{\emph{References/Bibliography}}}}
References or bibliographies are sources consulted.  Each
reference should contain name(s) of author(s), title of the paper,
journal name, volume number of the issue in which the article
appeared, starting page number, end page number and year of
publication.  In the case of a book source  its author(s), title,
publishers's name, place of publication, year of publication and
edition should be given.  Some examples are given below.
\renewcommand{\labelenumi}{(\theenumi)}
\renewcommand{\theenumi}{\arabic{enumi}}
\begin{enumerate}
 \item
Suppose the reference is the paper of K.~Murali, Sudeshna Sinha
and W.L.~Ditto with title ``Implementation of NOR gate by a
chaotic Chua's circuit" appeared in the journal called
`International Journal of Bifurcations and Chaos' in the year
2003,  the volume number of corresponding issue is 13 and the
starting and ending page numbers of the article are 2669 and 2672
respectively.  The above article can be specified as (without
mentioning the title of the article)
 \vskip 3pt
K.~Murali, Sudeshna Sinha and W.L.~Ditto, Int. J Bifur. and Chaos
13 (2003) 2669--2672.
\item
For an article which appeared in a conference proceedings a
typical format is given below:
 \vskip 3pt
R.~Harish and K.P.N.~Murthy, {\emph{Intermittency and multifractality
in iterated function systems}}. In: Nonlinear Systems. Eds.
R.~Sahadevan and M.~Lakshmanan (Narosa, New Delhi, 2002)
pp.~361--371.
 \vskip 5pt
In the above {\emph{Intermittency....}} is the title of the report of
R.~Harish and K.P.N.~Murthy. {\emph{Nonlinear Systems}} is the title of
the conference proceedings edited by R.~Sahadevan and
M.~Lakshmanan.  The proceeding was published in the year 2002 by
Narosa Publishing House, New Delhi.  In the proceedings the
article appears from the page 361 to page 371.
\item
A book can be noted down as, for example
 \vskip 3pt
T.~Kapitaniak, {\emph{Controlling Chaos}} (Academic Press, San Diego,
1996).
\item
A Ph.D. thesis can be referred as shown below:
 \vskip 3pt
S.~Parthasarathy, {\emph{On the analytic structure and chaotic dynamics
of certain damped driven nonlinear oscillators}}. Ph.D. thesis.
(Bharathidasan University, 1993, Unpublished).
\item
For an unpublished manuscript downloaded from internet one can
note down the web site where it is available (see for example the
references 7 and 8 of the references section of this manuscript).
\end{enumerate}
References can be either in alphabetical order according to
author's name or the order in which they are referred in the
report. Make sure that each reference cited in the text is
correctly entered into the list of references.  Repetition of
references in the list should be avoided.
\subsection{Typing the Report}
Typing should conform to the set of requirements of the
institution. The thesis should be double line spaced and not more
than 25 lines per page.  It may be typed on both sides. Chapter
heading must be in large size with bold face.  Each paragraph
should be right margin aligned.  Important terms when used first
time can be in italic letters and bold face.  First word of a
sentence should not be an abbreviation.  Latest softwares such as
LATEX or WORD can be used for thesis, dissertation and  report
preparation. One could download the software  LATEX a free of cost
from the web sites:
 \vskip 5pt
 1) http://www.ctan.org
 \vskip 1pt
 2) http://www.miktex.org
 \vskip 10pt
If a report is prepared keeping all the above precautions in mind,
there is every likelihood of it becoming useful for proper study.
Such report enables the reader to comprehend the data and to
determine for himself the validity of the conclusion.
 \vskip 10pt
{\emph{Before or immediately after submitting hard copies of the
Ph.D. dissertation to a university, show it to your colleagues,
teachers, scientists of your department, your parents and
friends}}.
\section{ACKNOWLEDGMENT}
We acknowledge valuable discussion with Professors M.~Sivasankaran
Nair, K.~Balasubramanian and E.~Subramanian.  We are very
grateful to Professors P.R.~Subramanian and K.P.N.~Murthy for a
critical reading of the manuscript and their suggestions which
greatly improved the presentation of the manuscript.  We are
thankful to Professors V.~Devanathan,  K.P.N.~Murthy and Sudeshna
Sinha for their suggestions to young researchers.
 \vskip 20pt
 \hrule
  \hrule
  \vskip 10pt
\noindent{\bf{REFERENCES:}}
\renewcommand{\labelenumi}{\theenumi.}
\renewcommand{\theenumi}{\arabic{enumi}}
\begin{enumerate}
\item
%1
M.R.~Beasley and L.W.~Jones, Physics Today June 1986 pp.36.
 \label{beas}
\item
%2
Research methodology in Yenza, http:// www.nrf.ac.za/yenza/research/internet.htm
\label{internet1}
\item
%4
C.R.~Kothari, {\emph{Research Methodology: Methods and
Techniques}} (Wiley Eastern, New Delhi, 1985).
 \label{kothari}
 \item
%5
P.~Saravanavel, {\emph{Research Methodology}} (Kitab Mahal,
Allahabad, 1987).
 \label{sara}
\item
%6
E.M.~Phillips and D.S.~Pugh, {\emph{How to get a Ph.D.?}}
(UBSPD, New Delhi, 1993).
 \label{phil}
\item
%7
R.~Spangenburg and D.K.~Moser, {\emph{The History of Science in
the Eighteenth Century}} (University Press, Hyderabad, 1999)
 \label{span}
\item
%8
B.S.~Warrier, The Hindu, 18 September 2006 pp.6 of Education Plus;
30 October 2006 pp.6 of Education Plus; 6 November 2006 pp.6 of 
Education Plus; 20 November 2006 pp.2 of 
Education Plus; 27 November 2006 pp.4 of 
Education Plus; 
4 December 2006 pp.6 of Education Plus.
\label{warrier}
\item
%9
Emotional factors in, http://www.cs.indiana. edu/mit.research.how.to/section3.13.html
 \label{cs}
\item
%10
Common errors made in research in,
http://sociology.camden.rutgers.edu/ jfm/tutorial/errors.htm
 \label{camden}
\item
%11
B.S.~Warrier, The Hindu, 11 May 2004.
 \label{warpre}
\end{enumerate}
 \vskip 10pt
 \hrule
  \hrule
 \vskip 20pt
\section*{Some quotations on Research}
 \vskip 10pt
\noindent {\emph{Research is what I'm doing when I don't know what
I'm doing}}. -- von Neuman
 \vskip 10pt
\noindent After the discovery of X-rays by R\"ontgen a journalist
interviewed him.
 \vskip 5pt
Journalist:  {\emph{What did you think?}}

R\"ontgen:  I didn't think, I investigated.

Journaist:  {\emph{What is it?}}

R\"ontgen:  I don't know.
 \vskip 10pt
\noindent {\emph{Research is key to our long-term position}}. -- Bill Gates
 \vskip 10pt
\noindent {\emph{It appears at first incredible that any discovery should be made, and when it has been made, it appears incredible that it should so long have escaped men's research.  All of which affords good reason for hope that a vast mass of inventions yet remains.}} -- Francis Bacon
 \vskip 10pt
\noindent Enrico Fermi was asked what characteristics physics
Nobelists had in common.  He answered, ``{\emph{I cannot think of
a single one, not even intelligence.}}"
 \vskip 10pt
\noindent {\emph{All progress is born of inquiry.  Doubt is often better than over confidence, for it leads to inquiry, and inquiry leads to invention.}}--Hudson Maxim
 \vskip 10pt
\noindent {\emph{Whenever one of my students came to me with a
scientific project, I asked only one question:
``Will it bring you nearer to God?"}}--I.I.~Rabi
 \vskip 10pt
\noindent {\emph{Scientific discovery and scientific knowledge have been achieved only by those who have gone in pursuit of it without any practical purpose whatsoever in view}}--Max Planck
 \vskip 10pt
\noindent {\emph{My success if you could call it that, lies in 
the fact that I have kept at my work all these years.  It is not
genius or anything, like that, it is merly patience}}. -- Annie Jump Cannon
 \vskip 10pt
\noindent {\emph{It seems to me that scientific research should be
regarded as a painter regards his art, a poet his poems, and a
composer his music}}. -- Albert A. Michelson
 \vskip 10pt
\noindent{\emph{Failing to plan is planning to fail}}. -- Allen
Lakein
 \vskip 10pt
\noindent {\emph{The average Ph.D. thesis is nothing but
transference of bones from one graveyard to another}}. -- Frank J.
Dobie
 \vskip 10pt
\noindent {\emph{When I got by B.S., I would be able to
bullshit... When I got by M.S. I would  have more shit, and that
finally, upon reaching my Ph.D., it would be piled higher and
deeper}}. -- S. Baker
 \vskip 10pt
\noindent {\emph{Works are of value only if they give rise to better ones}}.
 -- Alexander von Humboldt
 \vskip 10pt
\noindent {\emph{A hypothesis or theory is clear, decisive, and
positive, but it is believed by no one but the man who created it.
Experimental findings, on the other hand, are messy, inexact
things, which are believed by everyone except the man who did the
work}}. -- Harlow Shapley
 \vskip 10pt
\noindent {\emph{I keep six serving men,}} \\
\noindent {\emph{They taught me all I knew;}} \\
\noindent {\emph{Their names are What and Why and When}}, \\
\noindent {\emph{And Where and How and Who.}} 
 -- Rudyard Kipling
 \vskip 10pt
\noindent {\emph{The difficulty of literature is not to write , 
but to write what you mean. -- R.L.~Stevenson}}
\vskip 10pt
\noindent {\emph{I always preferred to try to imagine new
possibilities rather than merely to follow specific lines of reasoning
or make concrete calculations.  Some have this trait to a greater
extent than others.  But imagining new possibilities is more trying
than pursuing calculations and cannot be continued for too long a
time}}--S.M.~Ulam
\vskip 10pt
\noindent{\emph{My own research life has been greatly enriched by
having been broken into by periods of enforced change.  I was not
idle while I had my three children; far from it.  But it gave me
the opportunity of standing back, as it were, and looking at my
work.  And I came back with new ideas}}. -- Kathleen Lonsdale
 \vskip 10pt
\noindent My advice to young women students: {\emph{Don't quit. Muddle
through.  Get your 'union card' (Ph.D.) if you want to do
research.  Don't think you can't succeed if you're not first in
your class, or even in the middle; or even below that.  You will
increase your confidence as you go along ...}} -- Vera C. Rubin
 \vskip 10pt
\noindent My advice to young women scientists:  {\emph{To persevere, to love work and to love to do good work, to be independent, to be scientifically honest, and to embrace your ambitions, all the while respecting culture and responsibility to your family.  Knowledge and know-how are the way of liberty and equality.  Neither gender, nor religion, nor age will stand as a barrier to research.}} -- 
Zohra Ben Lakhdar
 \vskip 10pt
\noindent  {\emph{Research needs an inquisitive mind which is never satisfied with the current solution or state of affairs}} -- R.~Biswas
 \vskip 10pt
\noindent  {\emph{An age when the pizza delivery companies promise you a free dinner if they take more than half an hour to deliver is counter to the mindset needed for research}}. -- R.~Biswas
 \vskip 10pt
\noindent  {\emph{In science, self-satisfaction is death.  Personal self-satisfaction is the death of the scientist.  Collective self-satisfaction is the death of research.  It is the restlessness, anxiety, dissatisfaction  and agony of mind that nourish science.  }}. --R.~Jeyaraman
\vskip 50pt
%
%
%\newpage
{\noindent{\large{\emph{\bf{A Short interview with three eminent
scientists.}}}}}
 \vskip 10pt
\noindent{\bf{1. Interview with Professor V.~Devanathan}}
 \vskip 10pt
\noindent{\emph{What are the requirements for a successful
research career?}}
 \vskip 5pt
\noindent{\emph{Prof.~V.~Devanathan}} : Motivation and innate
interest in the topic of his research pursuit are the requirements
for a successful research career.  If a person takes the research
not by compulsion but by his own choice, then he will not feel it
as a burden but pursue it as a hobby.  {\emph{Science is at its
best when it is a part of a way of life}} - this is the
inscription that is found on the foundation stone of Institute of
Mathematical Sciences, Chennai and truly describes the correct
aptitude for a successful research career.
 \vskip 10pt
\noindent{\emph{Is it possible for an average student to come up
with novel results in a research problem?  If so, what kind of
approach he should follow?}}
 \vskip 5pt
\noindent{\emph{Prof.~V.~Devanathan}} :  Usually, the assessment
of a student as good, average or bad is based on his performance
in the examinations.  There are some who are good in examinations
with a good memory for reproduction but lack in deeper
understanding of the subject and originality in approach.  There
are some who are not so good in examinations but show originality
in thinking and follow unconventional or novel approach to the
subject.  There are a few who are good both in examinations and
research.  So, an average student with an ability of average
performance in the examinations, need not feel different if he has
{\emph{originality in thinking}} and {\emph{self-confidence}}.
 \vskip 10pt
\noindent{\emph{During a research career, a young researcher may
come across disappointing moments like not getting expected
results, rejection of a research article from a journal, etc. What
kind of mode of approach a researcher should have to face such
situations?}}
 \vskip 5pt
\noindent{\emph{Prof.~V.~Devanathan}} :  {\emph{Success begets success
and failure begets failure}}.  Success and failure are like two
sides of a coin and one is bound to face them alternatively in the
course of one's research career.  Elation at the time of success
and depression at the time of failure are usually mitigated if one
works in collaboration with others.  At the time of depression,
the co-workers come to the rescue and prop up the sagging spirit.
 \vskip 10pt
\noindent{\emph{In our manuscript we have mentioned the
following:}}
 \vskip 2pt
\noindent{\emph{Each and every bit of work has to be done by the
researcher. A young researcher should not do the entire work in
collaboration with others.  The researcher is advised to perform
all the work starting from identification of the problem to report
preparation by himself under the guidance of supervisor.}}
 \vskip 2pt
\noindent{\emph{Please give your views on this point.}}
 \vskip 5pt
\noindent{\emph{Prof.~V.~Devanathan}} : At the initial stages, the
researcher gets the support of the research group in which he is
working and he acquires the knowledge of the group effortlessly.
The weekly informal seminars, if conducted within the group, will
increase the pace of learning and help to clarify and crystallize
the problems.  This process of learning is made easier if the
young researcher works in collaboration with others.  This is true
both for theoretical and experimental work.  At present, the
experimental work is almost a team work and successful research
group is one in which the group leader allots the specified work
to individuals taking into account his ability and expertise.
 \vskip 10pt

\noindent{\bf{2. Interview with Prof~K.P.N.~Murthy}}
 \vskip 10pt
\noindent{\emph{The common belief is that research is laborious
and painful. Many times you have mentioned: {\emph{Doing research is an
entertainment}}.  Please, elaborate on this statement of yours.}}
 \vskip 5pt
\noindent{\emph{Prof~K.P.N.~Murthy}} :   Research not only
constitute a discovery or creating a new paradise but also consist
of obtaining a personalized understanding of a phenomenon.  The
struggle that you go through for obtaining an insight into a
phenomenon or getting a hold of a nuance and the extessy that you
get when you get an understanding of a phenomenon or obtaining a
new way of explaining of that phenomenon may be unmatched.  This
ecstasy is nothing to do with what yours creative have impact on
science and society. However, it is the ecstasy of what Einstein
got when he created special theory of relativity or Feynman when
he created quantum electrodynamics or Raman when he found the
so-called Raman lines. It is this makes the research an enterprise
of joy.  It is that makes a research an entertainment.
 \vskip 10pt
\noindent{\emph{Is it necessary for a beginner of research to
learn all the aspects of theoretical, experimental and numerical
techniques involved in a topic before he take-up an actual
research problem? }}
  \vskip 5pt
\noindent{\emph{Prof~K.P.N.~Murthy}} :  A certain basic knowledge
about physics and mathematics is must for starting research.
That is it.  Several things you learn doing research.  Ignorance
of even some of the basic elements is no hindrance for creativity.
What is required for doing good research is an enthusiasm, a
commitment and willingness to go back to basics and learn them right.
\vskip 10pt
\noindent{\emph{Before preparing the final write-up of your
research work, you have the practice of discussing the salient
features of your findings with a few other researchers. How are
you benefited from this?}}
 \vskip 5pt
\noindent{\emph{Prof~K.P.N.~Murthy}} :  After you have completed a
piece of work I find it is a good practice to discuss with your
colleagues the important findings that you have made.  I have
always realized that I got a better understanding of what I have
done when I tried to explain to my colleagues about my work in a
convincing way.  The very act of speaking of what you have done
removes the cob-webs in your understandings.  I always make it to
give a seminar on my work to a larger audience before submitting
it to a journal for publication.  I feel this is a very good and
helpful practice.
 \vskip 10pt
\noindent{\emph{Enjoy doing research and approach it as an
entertainment and a mode of getting happiness.  This is your
suggestion to young researchers. Please, brief it for the benefit
of youngsters. In what way will this be helpful to a researcher?
}}
 \vskip 5pt
\noindent{\emph{Prof~K.P.N.~Murthy}} :  In any human enterprise it is
important that one likes what one does.  The hard work that you have
put in a problem does not tired you and rest be assured if you approach
a research problem with joy and you will get a good result.
Publication of that result and the acceptance that you get from your
colleagues become secondary.  The satisfaction that you obtained by
doing a job well is a reward by itself.  I would say that youngsters
should have this attitude towards whatever they do.
 \vskip 10pt

\noindent{\bf{3. Interview with Prof~Sudeshna Sinha}}
 \vskip 10pt
\noindent{\emph{Despite unavoidable tasks a woman of our country
has, you have become one of the leading scientists in theoretical
physics.  What are your advice and suggestions to young
researchers particularly to young women researchers?}}
 \vskip 5pt
\noindent{\emph{Prof~Sudeshna Sinha}} :  It is indeed somewhat
harder for women to concentrate on career planning - especially
when their children are young.  One will have to accept that
household tasks will always be there.  The hardest thing is not
really the number of hours of work one can put in - but the
{\emph{quality of concentration}} one can achieve.  Here
discipline comes in.  Since women will probably manage to get
fewer hours of academic work done every day - they need to really
plan the academic work they hope to achieve every single day.  So
it is most beneficial to discipline oneself into shutting off all
daily chores {\emph{from one's mind}} for some hours every day.
The point is to learn efficiency -- and to appreciate that one does
not have the benefit of unlimited time (as others will make
justifiable demands on your time -- like children).

Also women may find it hard to pursue academic work at certain
points in their life - but they must preserve the self-confidence
and will to return to academic after such times are over.  They
must realize that in 3--4 decades of working life -- a few years is
not a big deal.  They should not think that a break in career is
{\emph{irreversible}}.
  \vskip 10pt
\noindent{\emph{Publishing in reputed journals (like Physical
Review Letters) is a dream or prestige for many physicists.  What
are the secret of yours for regular publications in  reputed
journals?  What type of problems one has to take up for getting
published in top-level journals? }}
 \vskip 5pt
\noindent{\emph{Prof~Sudeshna Sinha}} :  With journals like Physical
Review Letters one must remember two things:  First, always try
and make a case of the general interest of your results.  The
commonest grounds for rejection is {\emph{lack of broad
interest}}.  This is very subjective of course, and being Indian
does not help.  But still, at the outset, one should make an
attractive statement of the general scope of one's work (that is,
try to answer this hypothetical question:  Why should someone not
doing research in this exact narrow sub-field be interested in
reading my paper).  Second point is persistence.  Take all
criticisms of the paper seriously (and don't reply needlessly
aggressively to the referees) and try to answer all the
criticisms.  Then resubmit, and {\emph{don't give up till the last
round!}}
  \vskip 10pt
\noindent{\emph{How could a beginner of research come up with
novel results?}}
 \vskip 5pt
\noindent{\emph{Prof~Sudeshna Sinha}} :  Well, I think coming  up
with {\emph{novel}} results is not entirely in one's hand.  There
is an element of good fortune here!  If the guide of the young
researcher can identify a problem that is technically easy to
tackle -- but whose results can be of considerable potential
interest -- then there is a good chance for the young researcher to
get a novel result.  But this is not in the hands of the young
researcher, and most often not in the hands of the guide either
(as it depends on the subject, timing etc.).

{\bf{In this matter I always tell my students: whether you get a
novel result tomorrow is a matter of luck, but in a career
spanning several decades, if you work steadily and think deeply
about the subject, it is almost assured that at some point or the
other, you will get a good idea which will lead to a novel
result!}}

 \vskip 10pt
\noindent{\emph{To get a deep insight into the topic or problem of
research, what are the ways a young researcher can follow? }}
 \vskip 5pt
\noindent{\emph{Prof~Sudeshna Sinha}} : One should not just
passively {\emph{read}} papers or books!  One should try to work
it all out in some detail.  While reading passively one feels one
has {\emph{understood}} -- but only when one is trying to solve
something does one gain any real understanding.  In fact it is a
great idea to look at the title and abstract of a paper, and then
ask oneself how one would have attempted to work on such a problem
and only then look at what the authors have done.

\end{document}